\documentclass[aps,prb,twocolumn,floats,epsfig]{revtex4}
\usepackage{amssymb}
\usepackage{amsbsy}
\usepackage{amsmath}
\usepackage{epsfig}
\usepackage{color}
\usepackage{float}
\usepackage[colorlinks,linktocpage,bookmarks=false,citecolor=blue,linkcolor=red,urlcolor=blue]{hyperref}

\newcommand{\ing}{\includegraphics}
\newcommand{\bib}{\bibitem}
\newcommand{\beq}{\begin{equation}}
\newcommand{\eeq}{\end{equation}}
\newcommand{\bea}{\begin{eqnarray}}
\newcommand{\eea}{\end{eqnarray}}

\newcommand{\bra}[1]{\langle#1|}
\newcommand{\ket}[1]{|#1\rangle}

\begin{document}

\title{Dynamical Transition for a class of integrable models coupled to a bath}

\author{Madhumita Sarkar and K. Sengupta}

\affiliation{School of Physical Sciences, Indian
Association for the Cultivation of Science, 2A and 2B Raja S. C.
Mullick Road, Jadavpur 700032, India}

\date{\today}

\begin{abstract}

We study the dynamics of correlation functions of a class of
$d-$dimensional integrable models coupled linearly to a fermionic or
bosonic bath in the presence of a periodic drive with a square pulse
protocol. It is well known that in the absence of the bath,
these models exhibit a dynamical phase transition; all correlators
decay to their steady state values as $n_0^{-(d+1)/2}$[$n_0^{-d/2}]$
above [below] a critical frequency $\omega_c$, where $n_0$ is the
number of drive cycles. We find that the presence of a linearly
coupled fermionic bath which maintains integrability of the system
preserves this transition. We provide a semi-analytic expression for
the evolution operator for this system and use it to provide a phase
diagram showing the different dynamical regimes as a function of the
system-bath coupling strength and the bath parameters. In contrast,
when such models are coupled to a bosonic bath which breaks
integrability of the model, we find exponential decay of the
correlators to their steady state. Our numerical analysis shows that
this exponential decay sets in above a critical number of drive
cycles $n_c$ which depends on the system-bath coupling strength and
the amplitude of perturbation. Below $n_c$, the system retains the
power-law behavior identical to that for the closed integrable
models and the dynamical transition survives. We discuss the
applicability of our results for interacting fermion systems and
discuss experiments which can test our theory.

\end{abstract}

\maketitle

\section{Introduction}
\label{intro}

The physics of driven quantum systems has been actively studied in
recent years \cite{rev1}. Out of these, periodically driven systems
host several phenomena that do not have any analog in their
aperiodic driven counterparts \cite{rev2}. For example, periodic
drives may lead to generation of quantum states with non-trivial
topology even when the corresponding ground state of the system is
topologically trivial \cite{toporef1}. In addition, such driven
systems may lead to novel steady states which are otherwise
inaccessible \cite{ss1}. Moreover, driven quantum systems can lead
to stable phases of quantum matter which have no counterparts in
absence of a drive; such phases may be classified based on their
symmetries \cite{sym1}. These system also exhibit the phenomenon of
dynamic freezing where the starting state of the driven system
displays a perfect overlap with itself at the end of one or multiple
drive periods \cite{df1}. More recently, it was found that periodic
drives may lead to weak ergodicity breaking behavior; the drive
frequency may be tuned to switch between regimes displaying
relatively quick thermalization consistent with eigenstate
thermalization hypothesis (ETH) and long-time coherent oscillatory
dynamics which constitutes example of violation of ETH in
non-integrable systems without disorder \cite{rydref1}.

Such driven system also display the phenomenon of dynamical
transitions which can be thought as the non-equilibrium counterparts
of quantum phase transitions \cite{dt1,dt2}. A class of such
transition manifest themselves through cusp-like singularities in
their Lochsmidt echo; the origin of such singularities have been
shown to be due to crossing of non-analyticities (or Fisher zeroes)
of the dynamic free energy of the driven system \cite{dt1}. Such
transitions do not lead to perceptible changes in properties of
local correlation functions. In contrast, the second class of
transitions which is known to occur in driven closed integrable
quantum systems, manifest themselves through the approach of the
local correlation functions to their steady state values \cite{dt2}.
For a $d-$dimensional closed integrable model after $n_0 \gg 1$
cycles of the drive, the correlation functions are shown to decay to
their steady state values as $n_0^{-(d+2)/2}$ for high drive
frequencies and as $n_0^{-d/2}$ for low drive frequencies. These two
dynamical regimes are separated by a critical drive frequency
$\omega_c$ at which the transition occurs; indeed, for $d=1$ models,
it was shown that there could be several reentrant transition
between these two regimes. The reason for this transition was
analyzed in terms of Floquet Hamiltonian of such driven systems. It
was shown that such transition occur due to appearance of additional
extrema in the Floquet spectrum as the drive frequency is lowered
\cite{dt2,dt3}; in this sense, this phenomenon is analogous to first
order phase transitions in equilibrium statistical mechanics where
transitions occur due to appearance of additional minima in the
system's free energy. However, such transition have been shown to
exist for closed integrable models only; the fate of such transition
in either open or interacting quantum systems where the system can
be non-integrable has not been studied so far.

In this work, we study a class of $d-$ dimensional periodically
driven integrable quantum systems coupled to a fermionic or bosonic
bath focussing on the fate of such dynamical transitions in the
presence of these bath. These models describes a large class of spin
and fermion models such as Ising model in $d=1$, Kitaev model in
$d=2$, superconductors and  charge/spin density waves (CDW/SDW)
systems, and Dirac or Weyl like quasiparticles in graphene,
topological insulators (TI) and Weyl semi-metals (WSM). All these
systems are described by fermionic Hamiltonian given by
\begin{eqnarray}
H_0(t) &=& \sum_{\vec k} \psi^{\dagger}_{\vec k} \left[ (g(t)-
z_{\vec k} ) \tau_3 + \Delta_{\vec k} \tau_+ +{\rm h.c.} \right]
\psi_{\vec k} \label{intham}
\end{eqnarray}
where $\psi_{\vec k}= (c_{\vec k}, c_{-\vec k}^{\dagger})^T$ is a
two component fermionic field, $c_{\vec k}$ denotes fermion
annihilation operator, the sum over momenta extends over half of the
Brillouin zone, $\tau_{1,2,3}$ denotes Pauli matrices in the
particle-hole space, and the specific forms of $g(t)$, $z_{\vec k}$
and $\Delta_{\vec k}$ depend on the context of the model studied.
For example for the 1D Ising model, $g(t)$ denotes the transverse
magnetic field in units of the nearest neighbor interaction $J$
between Ising spins, $z_k=\cos k$ and $\Delta_k= i \sin k$ (where
the lattice spacing $a$ is set to unity)\cite{subirbook}. For the 2D
Kitaev model on a square lattice,depicting $p$-wave superconductors,
with parameters $J_{1,2,3}$ (where $J_3$ is chemical potential for
the fermions and $J_1$ and $J_2$ are their hopping strength and
pairing amplitude respectively) and unit lattice spacing, $z_{\vec
k} = (\cos(k_x) + \cos(k_y))$, $g(t)= J_3(t)/J_1$, and $\Delta_{\vec
k} = i J_2[\sin(k_x) + \sin(k_y)]/J_1$ \cite{kitref1}. This model is
topologically equivalent to the Kitaev spin model on the honeycomb
or brickwall lattices for $J_1=J_2$ \cite{kitspinrefs}. We note that
for Dirac quasiparticles the two component wavefunction is given by
$\psi'_{\vec k} = (c_{\vec k \uparrow}, c_{\vec k \downarrow})^T$
where $\sigma=(\uparrow, \downarrow)$ denote spin (for TIs and WSMs)
or pseudospin (for graphene) indices; such a wavefunction can be
easily mapped to $\psi_{\vec k}$ using a particle-hole
transformation. A similar consideration holds for wavefunctions of
CDW and SDW systems. In what follows we shall study the dynamics of
such a model driven periodically by varying $g(t)$ and coupled to a
fermionic/bosonic bath. Our numerical results would use examples of
1D Ising and 2D square lattice Kitaev models for p-wave
superconductors; however qualitatively similar features are expected
to hold for TI, WSM, and graphene quasiparticles, Kitaev spin models
on honeycomb and brick-wall lattices, and CDW/SDW systems mentioned
above.

The main results that we find from such a study are as follows.
First, for $H_0$ coupled linearly to a non-interacting bath which
retains the integrability of the system, we obtain an exact
semi-analytic expression for the evolution operator $U$ and hence
the Floquet eigenspectrum. Using the properties of the Floquet
spectrum and also via explicit calculation of dynamic behavior of
system correlation functions, we show that the system displays
dynamical transitions. Second, our analysis finds that the
system-bath coupling strength can be tuned to induce additional
dynamical transitions for high drive frequencies where the closed
system always remains in the high frequency phase; such transitions
have no analog in closed driven systems studied earlier. We provide
a comprehensive phase diagram charting out the positions of
different dynamical phases as a function of the drive frequency and
the bath parameters. Third, for $H_0$ coupled to a bosonic bath
which destroys integrability of the system, we find that all
correlators, at long drive times, always decay exponentially to
their steady state values. Such an exponential decay of the
correlators is characterized by decay constants. We analyze these
driven systems by using an equation of motion approach
\cite{eomref1} and chart out the behavior of these decay constants
as a function of drive frequency. Fourth, we find that such
exponential decay of correlators sets in after a critical number of
drive cycles $n_c$; for $n_0 \ll n_c$, the correlation functions
display power law behavior similar to their closed counterpart. We
chart out the dependence of $n_c$ on the drive amplitude and the
system-bath coupling strength. Our analysis demonstrates that the
dynamical transition of the closed system survives till a large
number of drive cycles at weak system-bath coupling and low drive
amplitude. Finally, we discuss the applicability of our analysis to
weakly-interacting fermionic systems and chart out experiments which
can test our theory.

The plan for the rest of the paper is as follows. In Sec.\
\ref{fermion}, we discuss the dynamical transitions in the presence
of a fermionic bath. This is followed by Sec.\ \ref{boson} where we
chart out the fate of such transitions in the presence of a bosonic
bath. Finally, we discuss our main results, chart out experiments
which can test our theory, and conclude in Sec.\ \ref{diss}.

\section{Fermionic Bath}
\label{fermion}

In this section, we shall discuss the dynamics of the integrable
models described by $H_0$ (Eq.\ \ref{intham}) coupled to fermionic
bath. The properties of the Floquet spectrum of the system is
described in Sec.\ \ref{secfl} while a phase diagram indicating
different dynamical regimes is presented in Sec.\ \ref{secpd}.

\subsection{Floquet Hamiltonian}
\label{secfl}

The total Hamiltonian for the integrable model $H_0$ (Eq.\
\ref{intham}) linearly coupled to fermionic bath can be written as
\begin{eqnarray}
H &=& H_0(t) + H_{\rm int} + H_{b}\label{htot1}
\end{eqnarray}
where $H_{b}$ is the bath Hamiltonian, and the interaction between
the system and bath is modeled by $H_{\rm int}$. The periodic drive
is implemented via a square pulse drive protocol,
\begin{eqnarray}
g(t)&=& g_i, 0 \leq t < T/2 \nonumber  \\
&=& g_f, T/2 \leq t <T  \label{proto}
\end{eqnarray}
where $T=2\pi/\omega_D$ is the time-period of the drive and
$\omega_D$ is the drive frequency. For the 1D Ising model $g(t)$
indicates time varying magnetic field while for the Kitaev p-wave
model $g(t)=J_3(t)$. The bath Hamiltonian is given by
\begin{eqnarray}
H_b &=& \sum_{\vec k} \epsilon_{b}(\vec k) f_{\vec k}^{\dagger}
f_{\vec k} \label{bathham}
\end{eqnarray}
where $f_{\vec k}^{\dagger}$ is the creation operator for bath
fermions $\epsilon_{b}(\vec k) = \eta \sum_{i=1,d} \cos k_i$ where
$\eta$ is a constant. Such a bath Hamiltonian constitutes the
simplest possible choice of tight-binding non-interacting fermion
model; in this work, we shall restrict ourselves to this model for
concreteness. We choose the spatial dimension of the bath to be same
as that of the system; thus for the Ising chain we choose a 1D bath
with $\epsilon_{b}(k) = \eta \cos(k)$ while for the Kitaev model
$\epsilon_b(\vec k)= \eta (\cos(k_x) + \cos(k_y))$. The interaction
between the system and the bath is described by
\begin{eqnarray}
H_{\rm int}=\sum_{\vec k} \left(\lambda_{\vec k}  c_{\vec k} f_{\vec
k}^{\dagger} +{\rm h.c.}\right)
\end{eqnarray}
where $\lambda_{\vec k}$ is the coupling function. For numerical
studies on transverse field Ising chain or Kitaev model, we shall
take $\lambda_{\vec k}=\lambda$ to be a constant.

For the closed system, it is well known that all correlators exhibit
one or multiple dynamical transition(s) as a function of the drive
frequency; the critical frequency of this transition can be inferred
from the eigenspectrum of it's Floquet Hamiltonian \cite{dt2,dt3}.
Thus we compute the Floquet spectrum of the system described by
$H$(Eq.\ \ref{htot1}) and subjected to a periodic drive given by
Eq.\ \ref{proto}. In what follows, we shall use the path-integral
technique developed in Ref.\ \onlinecite{roop} for computation of
the Floquet Hamiltonian. In this method, one express the matrix
elements of the evolution operator $\hat U$ of a quantum many body
system between two coherent states in imaginary time at a
temperature $T_0$. This is followed by a Wick-rotation to real time
$\beta = 1/(k_B T_0) \to i T/ \hbar $, where $k_B$ is the Boltzmann
constant; such a rotation can be analytically done for driven
Gaussian system for the protocol given in Eq.\ \ref{proto}. This
allows one to obtain $\hat U$ analytically in real time; the form of
the Floquet Hamiltonian can then be read off from the expression of
$\hat U$. It was shown in Ref.\ \onlinecite{dt2} that this method
reproduce the exact Floquet Hamiltonian for closed integrable Dirac
systems whose Hamiltonians are given by $H_0(t)$.

We begin by computing the evolution operator for the system for the
square pulse protocol (Eq.\ \ref{proto}) which is given by
\begin{eqnarray}
\hat{U}(T,0)&=&\hat{U}(T,T/2)  \hat{U}(T/2,0) = \hat U_f  \hat U_i\nonumber\\
&=& e^{-i H[g_f]T/(2 \hbar)} e^{-i H[g_i] T/(2 \hbar)}
\label{evolop1}
\end{eqnarray}
To obtain the Floquet Hamiltonian we first compute the matrix
elements of $\hat{U}_f$ and  $\hat{U}_i$ between two arbitrary
coherent states. For this we note that the two component system
fields are either given by $\psi_{\vec k}= (c_{\vec k}, d_{\vec
k})^T$ (for CDW/SDW systems) or as $\psi_k= (c_k,
c_{-k}^{\dagger})^T$ (for Ising and Kitaev models and
superconductors); for the latter class, we shall follow Ref.\
\onlinecite{roop} and perform a particle-hole transformation
$c_{-k}^{\dagger} \to d_{k}$ so that one can have a uniform
formalism for both the cases. No such transformations were carried
out for the bath fields. Using this, and performing the Wick's
rotation mentioned above we get
\begin{eqnarray}
\langle{\Phi^1_{\vec k}}| \hat U_{\vec k a} |{\Phi^2_{\vec k}}
\rangle &=& \exp[-\Phi^{1 \ast}_{\vec k} \,{\mathcal L}_{\vec k
a}\, \Phi^2_{\vec k}] \nonumber\\
\Phi^{b \ast}_{\vec k} &=& (\psi_{\vec k}^{\ast}, \psi_{-\vec
k}^{\ast}, \psi^{' \ast}_{\vec k}, \psi_{-\vec k}^{' \ast} )
\end{eqnarray}
where $b=1,2$, $a=i,f$, $\psi_{\pm \vec k}$ and $\psi'_{\vec k}$
denotes fermionic coherent states for the system and bath
respectively and $\hat U_a = \prod_{\vec k > 0} \hat U_{\vec k a}$.
Here ${\mathcal L}$ can be written as
\begin{eqnarray}
{\mathcal L}_{\vec k a} &=& I-G_a^{-1}(\vec k,0^+), \nonumber\\
G_a(\vec k, 0^+) &=& \frac{1}{\beta} \sum_{\omega_n} G_a(\vec k,
i\omega_n) e^{-i \omega_n \eta} \label{lexp1}
\end{eqnarray}
where $\beta= 1/(k_B T)$ is the inverse temperature, $I$ denotes the
$4 \times 4$ identity matrix, $\omega_n$ denotes the Matsubara
frequency, the index $a$ takes value $a=i,f$, the limit $\eta \to
0^+$ is to be taken at the end of the calculation, and $G_{i(f)}$
denotes the Green function of the system corresponding to
$g=g_i(g_f)$ whose calculation shall be charted out later in this
section. Thus we obtain the matrix element of $\hat U= \hat U_f \hat
U_i$ as
\begin{eqnarray}
\bra{\Phi_{\ell}}\hat U\ket{\Phi_{\ell'}}&=&\bra{\Phi_{\ell}} \hat
U_f \hat U_i\ket{\Phi_{\ell'}}
\nonumber\\
&=& \int D \Phi' D \Phi^{' \ast} e^{-\sum_{\vec k} |\Phi'_{\vec
k}|^2 + \Phi_{\ell \vec k}^{\ast} {\mathcal L}_{\vec k f}
\Phi'_{\vec k} +
\Phi^{'\ast}_{\vec k} {\mathcal L}_{\vec k i} \Phi_{\ell' \vec k}}\nonumber\\
&=& \exp[-\sum_{\vec k} \Phi_{\ell \vec k}^{\ast} {\mathcal L}_{\vec
k f} {\mathcal L}_{\vec k i} \Phi_{\ell' \vec k}]  \label{flou1}
\end{eqnarray}
Since $\Phi_{\ell \vec k}^{\ast}$ and $\Phi_{\ell' \vec k}$ are
arbitrary coherent states, one can identify the evolution operator
as
\begin{eqnarray}
U_{\vec k}(T,0) &=& {\mathcal M}_{\vec k} = {\mathcal L}_{\vec k f}
{\mathcal L}_{\vec k i}. \label{uevol1}
\end{eqnarray}
In particular, the eigenvalues of ${\mathcal M}_{\vec k}$,
$\lambda_{\vec k n}$, are related to those of the Floquet
Hamiltonian, $\epsilon_{\vec k n}^F$ as \cite{roop}
\begin{eqnarray}
\lambda_{\vec k n}= \exp[-i \epsilon^F_{\vec k n} T/\hbar ]
\label{eigen1}.
\end{eqnarray}

Next, we chart out the computation of $G_{i(f)}(0^+)$. To this end,
we write the action corresponding to $H$ (after the particle-hole
transformation discussed earlier in the section) as \cite{book3}
\begin{eqnarray}
S[\Phi^{\ast},\Phi] &=& \int_0^{\beta} d\tau (\Phi^{\ast} I
\partial_{\tau} \Phi + H_a[\Phi^{\ast}, \Phi]) \label{ac1}
\end{eqnarray}
where $\Phi$ is the four component field and $H_a$ denotes the full
Hamiltonian (Eq.\ \ref{htot1}) with $g=g_a$ and $a=i,f$. Using Eq.\
\ref{ac1}, one obtains
\begin{widetext}
\begin{eqnarray}
G_a^{-1}(\vec k, \omega_n)=-\left(
\begin{array}{cccc}
i \omega -\epsilon[{\vec k};g_{a}]  & -\Delta_{\vec k}  & -\lambda_{\vec k}  & 0 \\
 -\Delta_{\vec k}  & \epsilon[\vec k;g_{a}] +i\omega  & 0 & -\lambda_{\vec k}  \\
 -\lambda^{\ast}_{\vec k}  & 0 & i\omega -\text{$\epsilon_{b}(\vec k) $} & 0 \\
 0 & -\lambda^{\ast}_{\vec k}  & 0 & \text{$\epsilon_{b}(\vec k) $}+i\omega  \\ \label{green1}
\end{array}
\right)
\end{eqnarray}
\end{widetext}
where $\epsilon[{\vec k};g_{a}]=g_a - z_{\vec k}$. Using Eq.\ \ref{green1}, it is easy to find $G_a(\vec
k,\omega_n)$. In particular we find that poles of these equations,
assuming $\lambda_{\vec k}$ to be real, are given by the solution of
the equation
\begin{eqnarray}
&&\omega_n^4 +\omega_n^2(\Delta_{\vec k}^2+\epsilon[\vec k;
g_a]^2+\epsilon^2_{b}(\vec k)+2 \lambda_{\vec k}^2)+ \Delta_{\vec
k}^2 \epsilon^2_{b}(\vec k) \nonumber\\
&& +\epsilon^2[\vec k; g_a] \epsilon_{b}^2(\vec k)+\lambda_{\vec
k}^4 - 2 \epsilon[\vec k; g_a] \epsilon_{b}(\vec k) \lambda_{\vec
k}^2=0 \label{poleeqn}
\end{eqnarray}
and are given by $\omega_{1..4} = \pm \sqrt{\alpha_{\vec k} \pm
\sqrt{\beta_{\vec k}}}$, where
\begin{eqnarray}
&&\alpha_{\vec k} = [\Delta_{\vec k}^2 +2 \lambda_{\vec
k}^2+\epsilon^2[\vec k;g_{a}]+ \epsilon^2_{b}(\vec k)]/2
\label{alphabetaexp} \\
&& \beta_{\vec k} =[\Delta_{\vec k} ^2 +2
\lambda_{\vec k}^2+\epsilon^2[\vec k,g_a]+\epsilon_{b}^2 (\vec k)]^2/4 \nonumber\\
&&-[\Delta_{\vec k}^2 \epsilon_{b}^2(\vec k) +\lambda_{\vec k} ^4
+\epsilon^2[\vec k; g_a] \epsilon_{b}^2(\vec k) -2 \lambda^2_{\vec
k} \epsilon[\vec k;g_a] \epsilon_{b}(\vec k)] \nonumber
\end{eqnarray}
Using Eq.\ \ref{poleeqn} one obtains
\begin{eqnarray}
G_{a}(\vec k,\omega_n) &=& \prod_{i=1,4} (i \omega_n -
\omega_i)^{-1} {\mathcal C}_a(\vec k,\omega_n)\label{gmatrix}
\end{eqnarray}
where ${\mathcal C}_a$ denotes the adjoint of the cofactor matrix of
$G_a^{-1}$. From Eq.\ \ref{gmatrix}, one can compute
\begin{eqnarray}
G_a(\vec k,0^+) = \sum_{i=1,4} \frac{[1-n_F(\omega_i)] {\mathcal
C}_a(\vec k,\omega_i)}{\prod_{j\neq i,j=1,4}(\omega_i - \omega_j)}
\label{greentau}
\end{eqnarray}
where $\omega_i$ are the poles of the Greens's function (Eq.\
\ref{gmatrix}) and $n_F(\omega_i)=(1+\exp[\beta \omega_i])^{-1}$ is
Fermi-Dirac distribution function. This allows us to obtain
expression for ${\mathcal L}_{\vec k a}= I-[G_a(\vec k,0^+)]^{-1}$
and subsequently ${\mathcal M}_{\vec k}$ using Eq.\ \ref{greentau}
and \ref{lexp1}. This leads to the Floquet eigenvalues
$\epsilon_{\vec k n}^F$ (Eq.\ \ref{eigen1}).

\begin{figure}[H]
\centering {\ing[width=0.48\linewidth
,height=0.37\columnwidth]{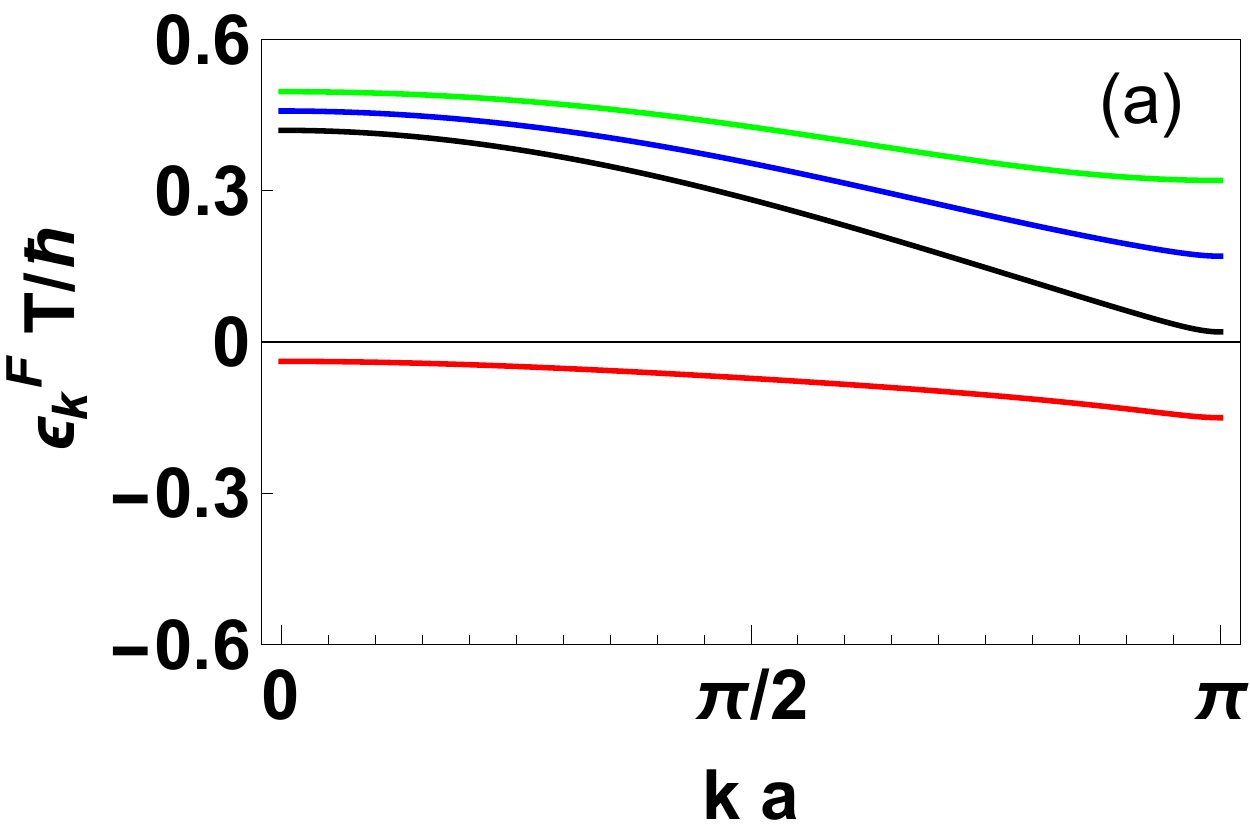}}\centering
{\ing[width=0.44\linewidth ,height=0.365\columnwidth]{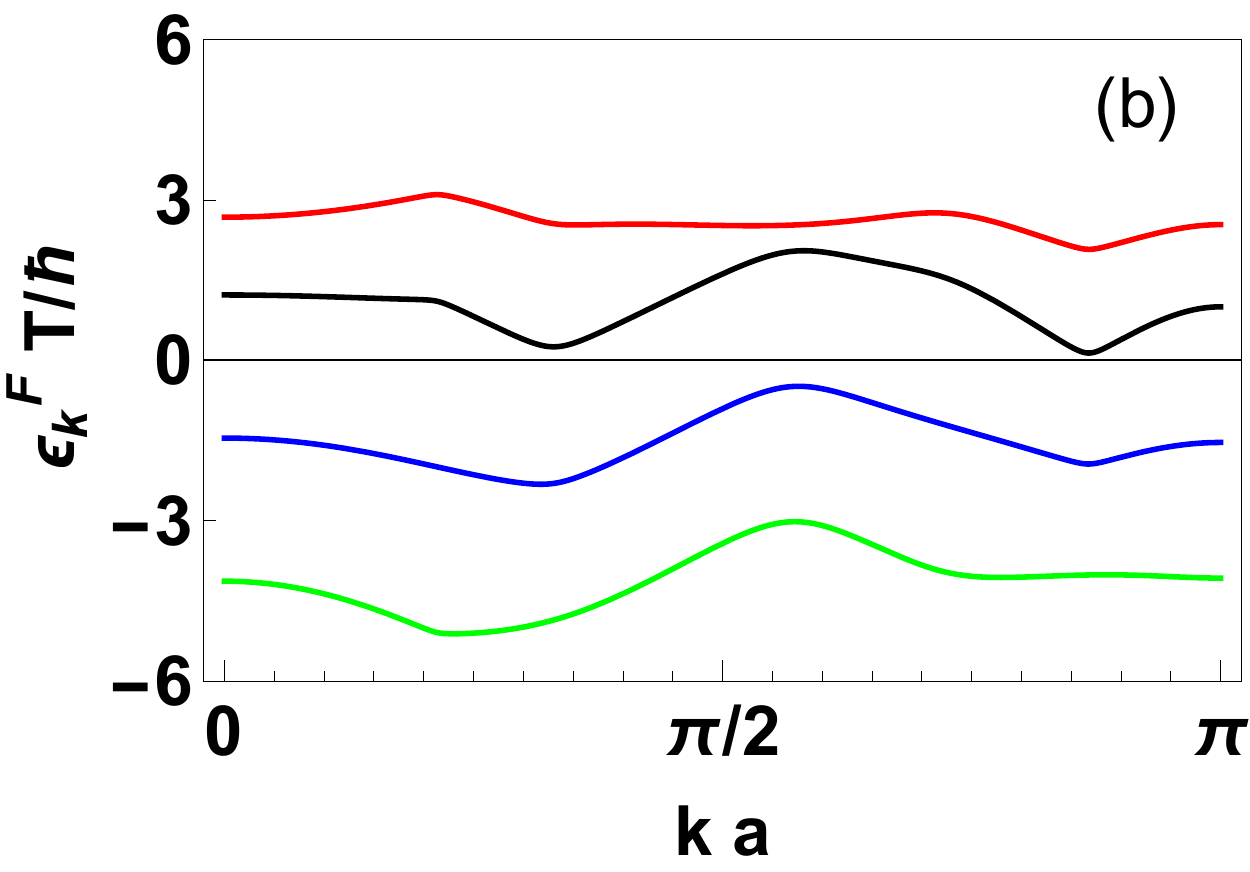}}
\caption{Plot of Floquet eigenvalues $\epsilon^F_{k,n} T/\hbar$ for
1D Ising model as a function of $k$ for $n=1..4$. (a)[(b)]
shows their dispersion at high [low] frequency
$\omega_D=10[0.2] \pi$. For all plots, $g_f=2$, $g_i=0$, and both
the lattice spacing $a$, and the Ising interaction strength $J$ is
set to unity.} \label{fig1}
\end{figure}

A plot of the Floquet eigenvalues for Ising model in a transverse
field is shown in Fig.\ \ref{fig1}. Here the lattice spacing $a$ and
the Ising interaction strength $J$ is set to unity, $z_{k}=\cos(k)$,
$\Delta_k= i \sin(k)$, $g_f/J=2$, $g_i/J=0$, and we have chosen
representative values $\lambda/J=0.8$ and $\eta/J=0.1$ for these
plots.  Fig.\ \ref{fig1}(a) shows $\epsilon_{k,n}^F T/\hbar$ as a
function of $k$ for $\omega_D=10 \pi$ while Fig.\ \ref{fig1}(b)
shows the corresponding plot at $\omega_D=0.2 \pi$. We find that at
low frequency Floquet eigenvalues display multiple extrema as shown
in the Fig.\ \ref{fig1}(b); this is in sharp contrast to their
behavior at high frequency shown in Fig.\ \ref{fig1}(a) where the
extrema are only found at $k=0,\pi$. This behavior indicates the
possibility of a dynamical transition at finite $\lambda$ and
$\eta$; this will be discussed in details in Sec.\ \ref{secpd}.

\begin{figure}[H]
\centering
\includegraphics[width=0.48 \linewidth]{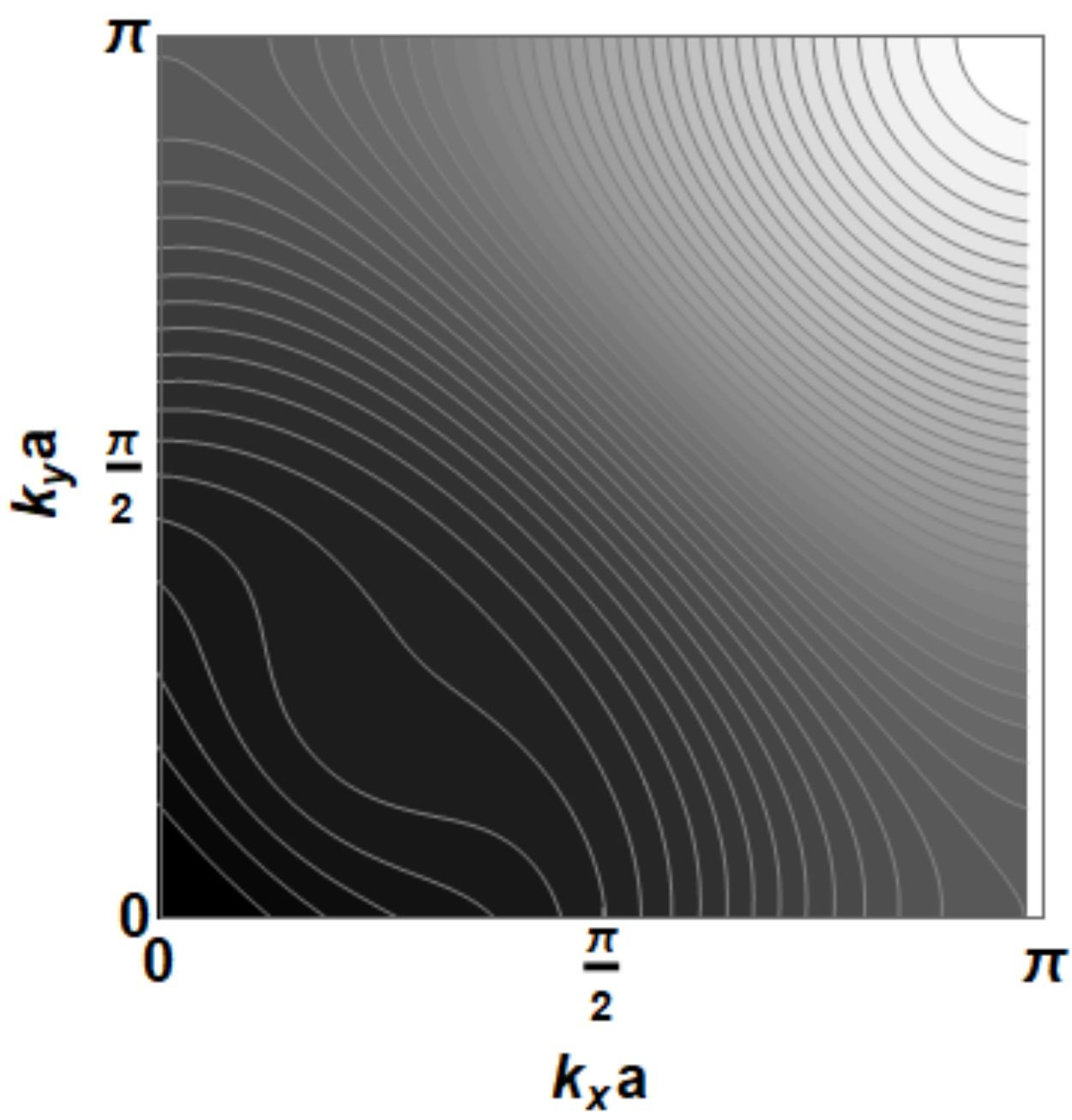}
\includegraphics[width=0.48 \linewidth]{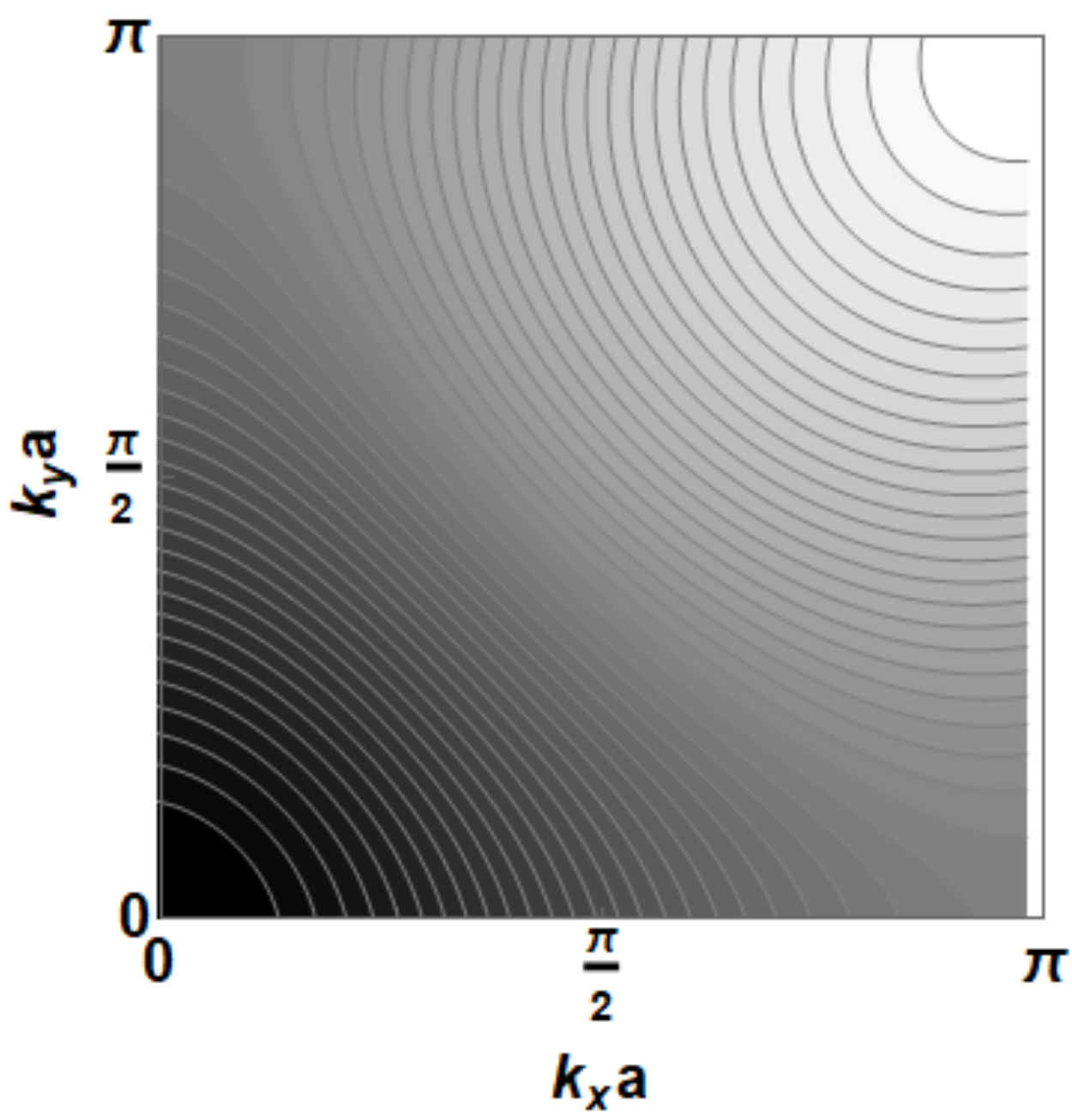}
\includegraphics[width=0.48 \linewidth]{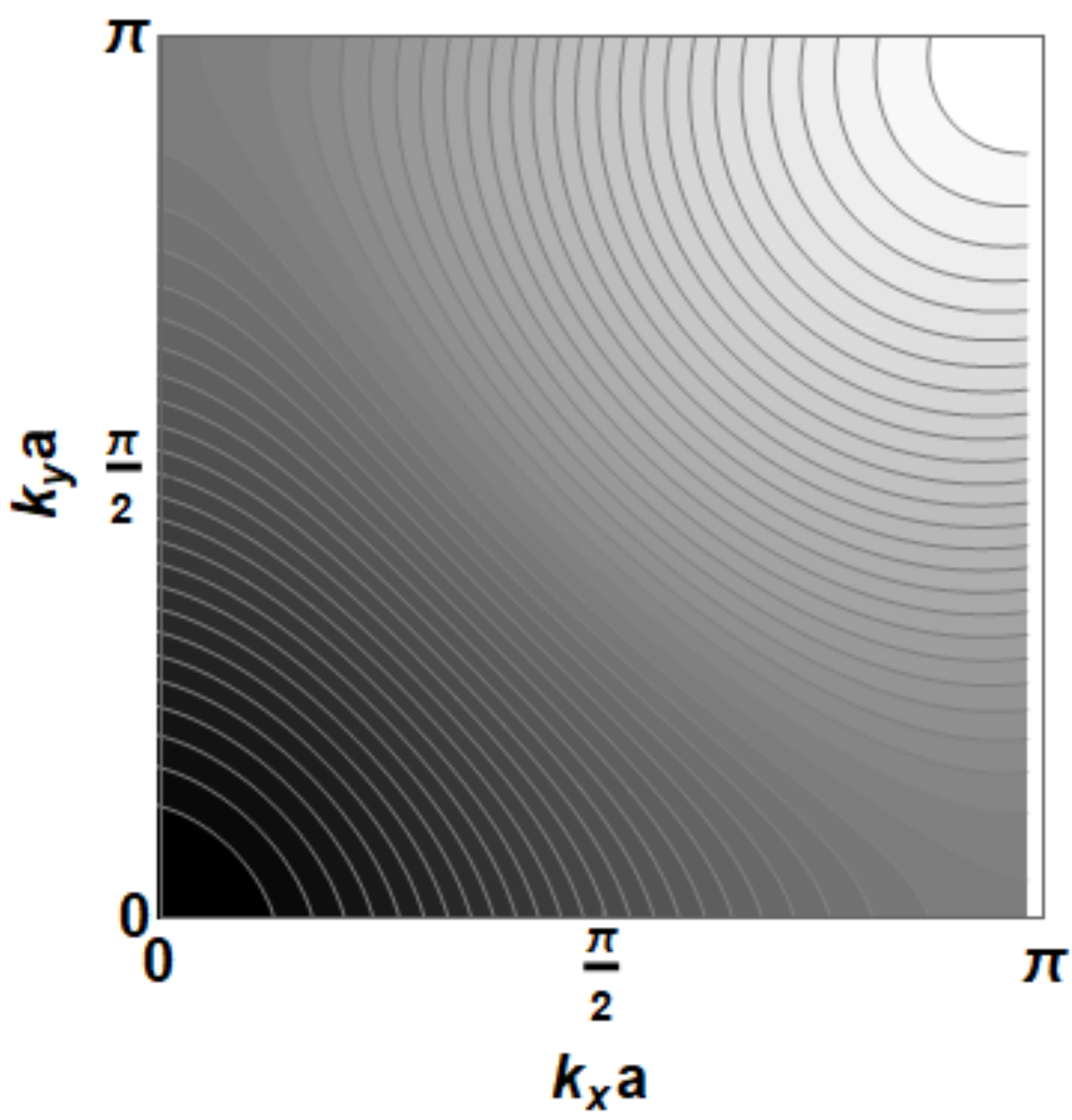}
\includegraphics[width=0.48 \linewidth]{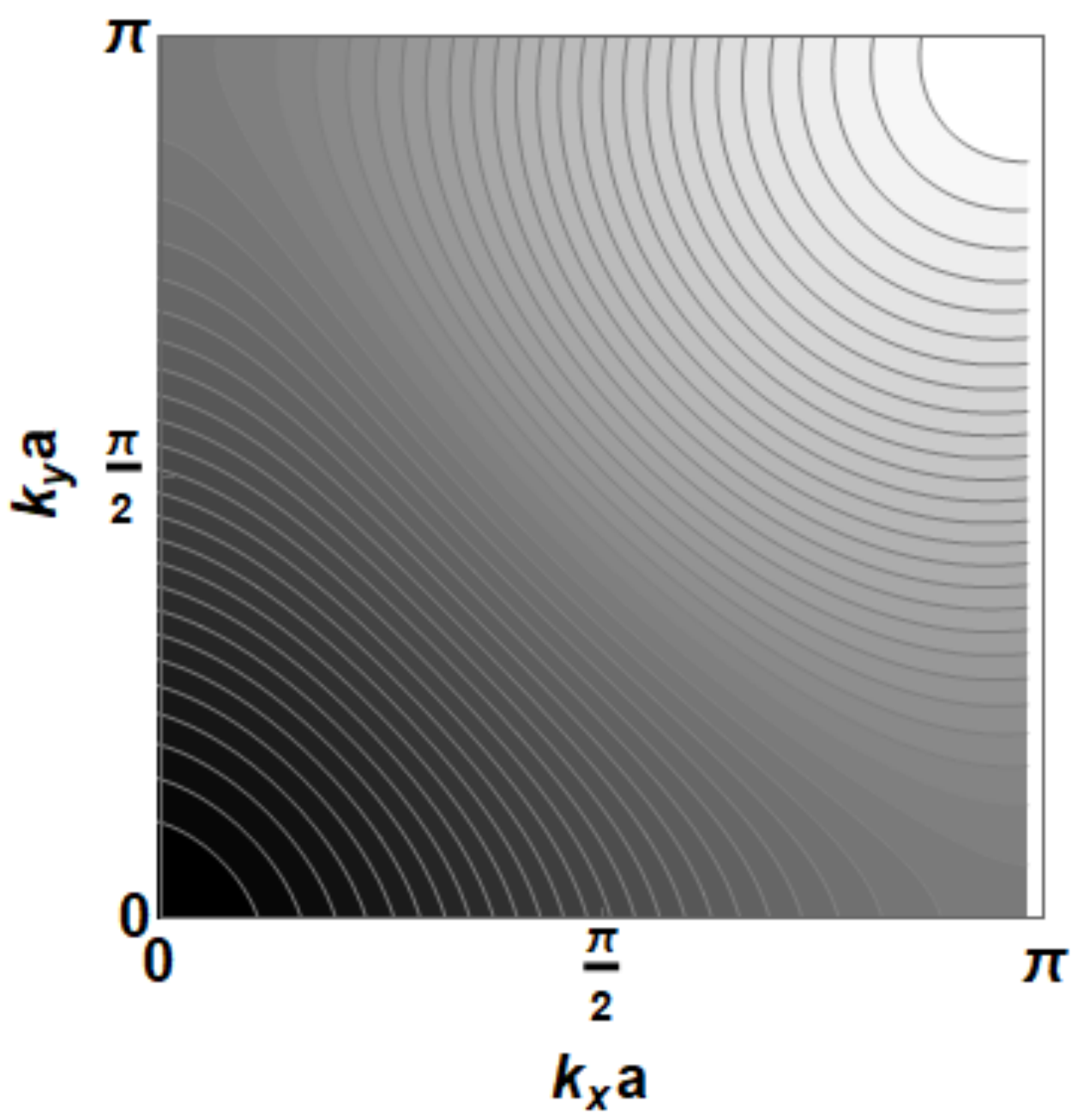}
\caption{Plot of Floquet eigenvalues $\epsilon^F_{\vec k} T/\hbar$
for 2D Kitaev model as a function of $\vec k=(k_x,k_y)$. Each panel
corresponds to one of the eigenvalues (four in total). For all
plots, $\hbar \omega_D/J_1=10\pi$, $g=J_3$ with $g_f=5J_1$ and
$g_i=4J_1$, $\lambda=0.8 J_1$, $\eta=0.1 J_1$, $J_2=J_1=1$ and the
lattice spacing $a$ is set to unity. The white (black) regions
denote high (low) values. See text for details.} \label{fig2}
\end{figure}

\begin{figure}[H]
\centering
\includegraphics[width=0.48 \linewidth]{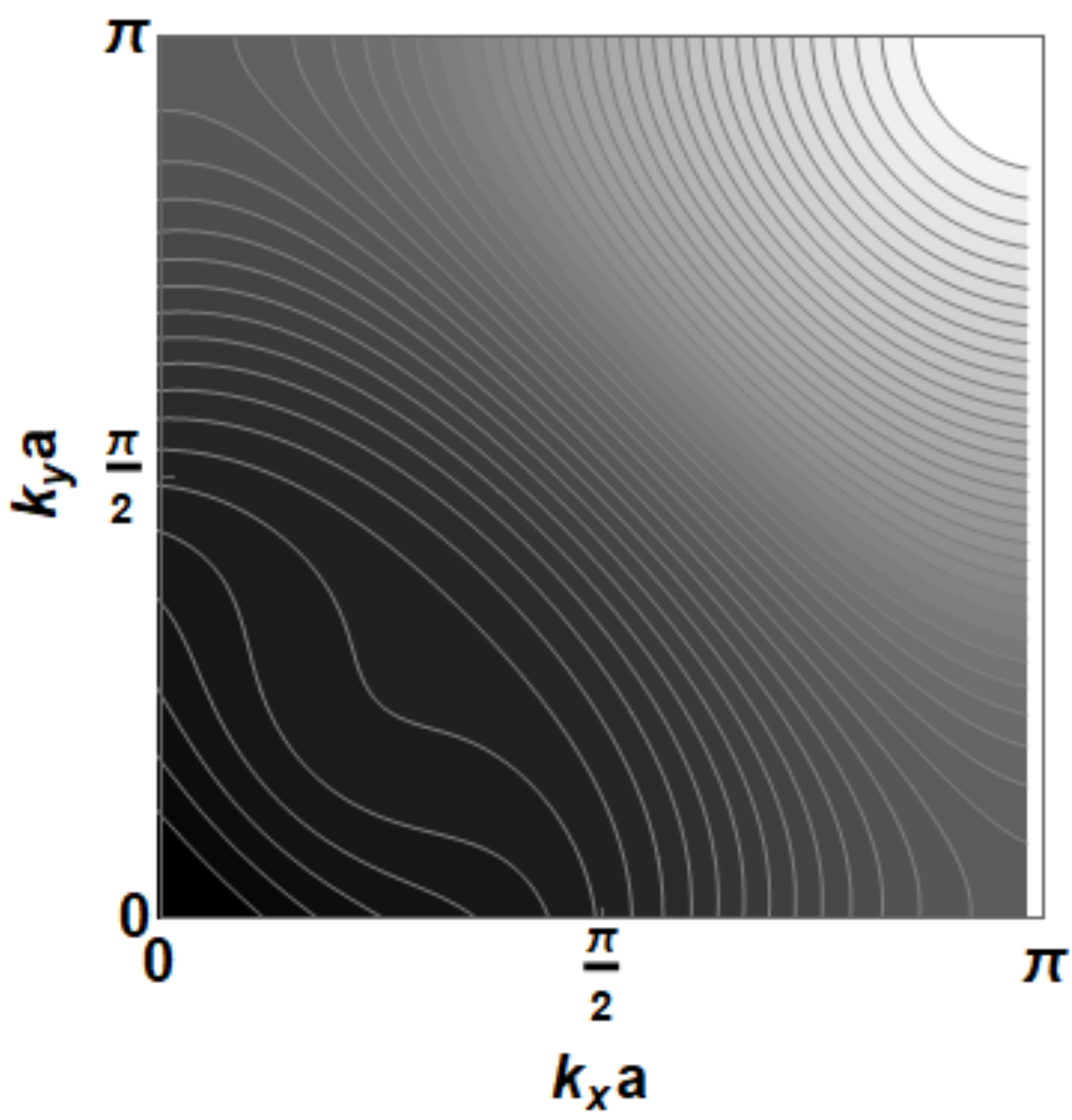}
\includegraphics[width=0.48 \linewidth]{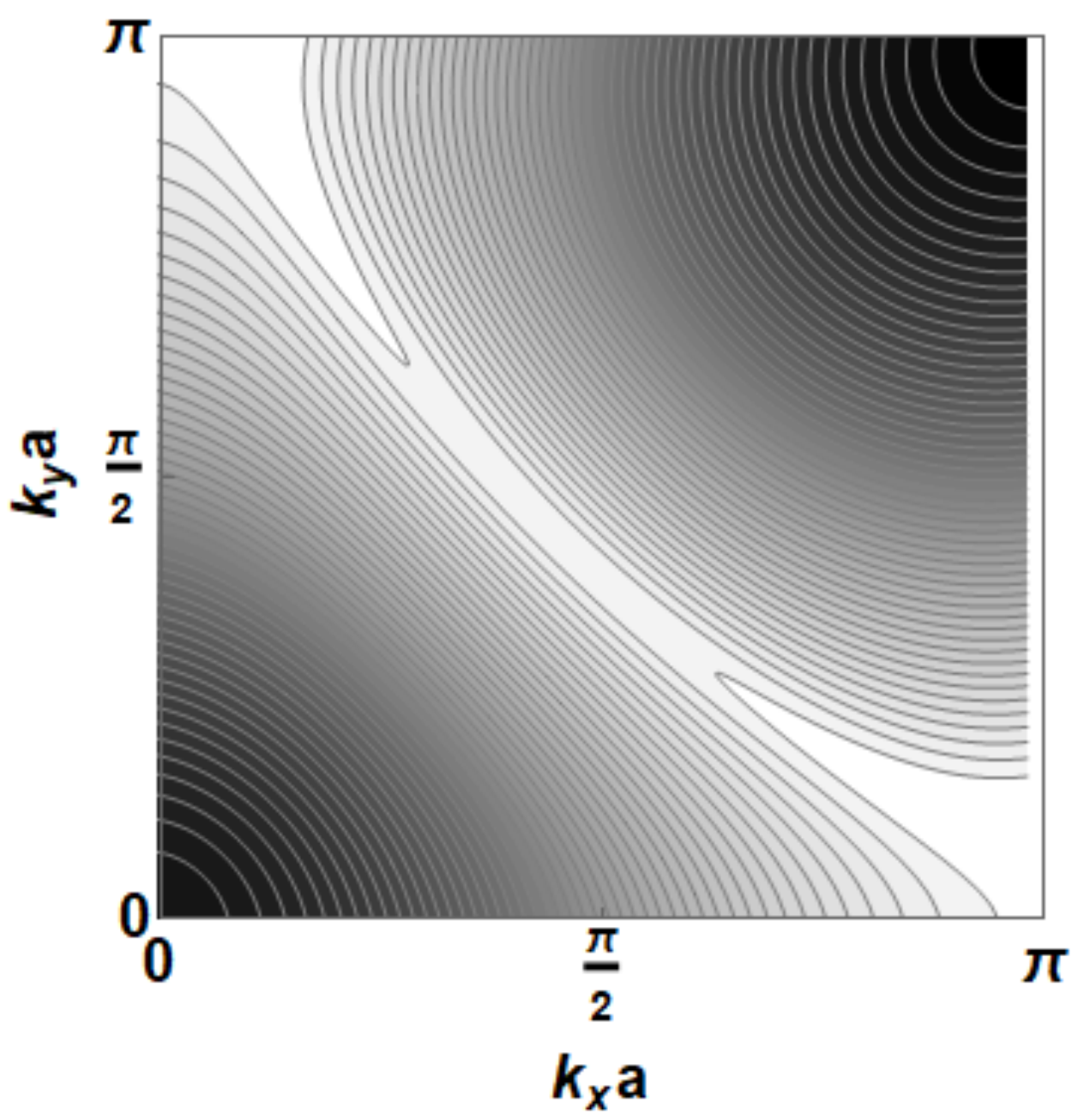}
\includegraphics[width=0.48 \linewidth]{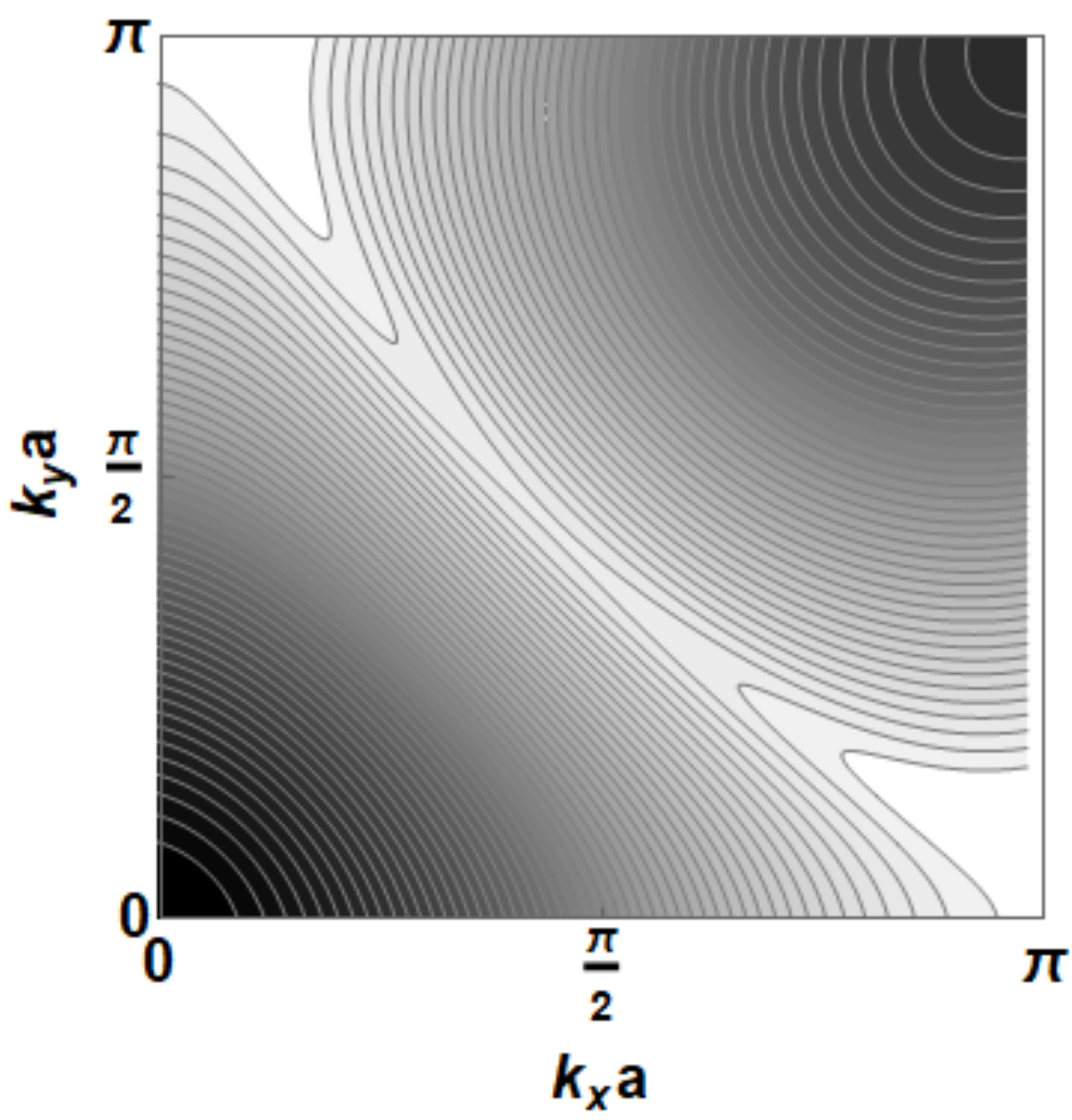}
\includegraphics[width=0.48 \linewidth]{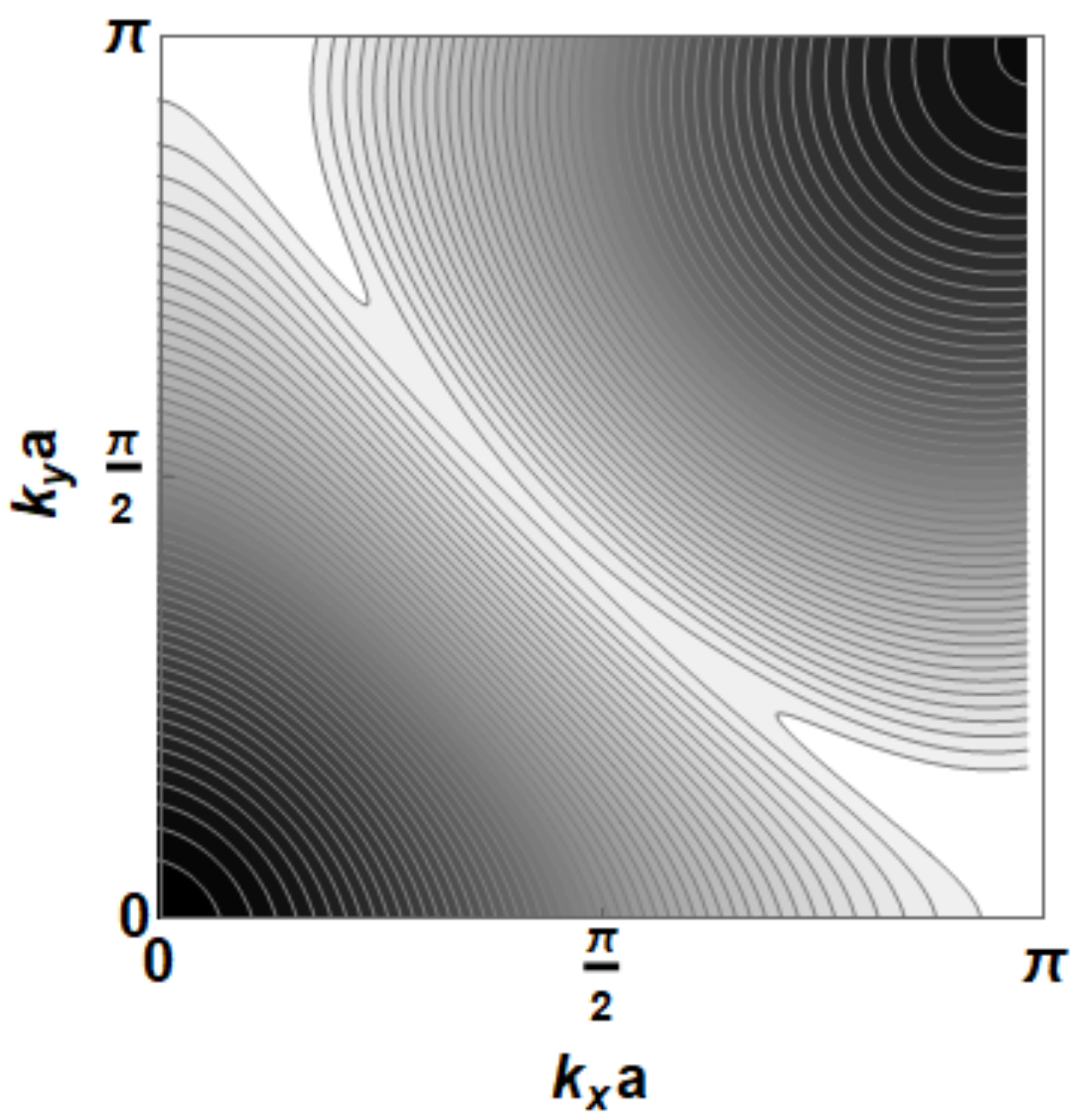}
\caption{Plot of Floquet eigenvalues $\epsilon^F_{\vec k} T/\hbar$
for 2D Kitaev model as a function of $\vec k=(k_x,k_y)$. Each panel
corresponds to one of the eigenvalues (four in total). For all
plots, $\hbar \omega_D/J_1=3.3 \pi$. All other parameters are same
as in Fig.\ \ref{fig2}. See text for details.} \label{fig3}
\end{figure}

Similar plots for the Floquet eigenvalues for the 2D Kitaev model in
the gapped phase is shown in Figs.\ \ref{fig2} and \ref{fig3}. For
the Kitaev model, $\Delta_{\vec k}= J_2 i( \sin(k_x a) + \sin(k_y
a))$, $z_{\vec k}= J_1 (\cos(k_x a) + \cos(k_y a))$, $J_3(t)/J_1=
g(t)$, and $a$ is the lattice spacing. Here, we have chosen
$J_1=J_2=1$, $J_{3f}=5J_1$, $J_{3i}=4J_1$, $\lambda=0.8 J_1$ and
$\eta=0.1 J_1$. Figs.\ \ref{fig2} and \ref{fig3} display four
Floquet eigenvalues at $\hbar \omega_D/J_1=10 \pi$ and
$\hbar\omega_D/J_1= 3.3 \pi$ respectively. We find that at high
frequency ($\hbar \omega_D/J_1=10 \pi$), all the Floquet eigenvalues
show extrema at the band edges or center (Fig.\ \ref{fig2}); in
contrast, one finds an arc of maxima for three of the four
eigenvalues at low frequency ($\hbar \omega_D/J_1= 0.2 \pi$). As we
shall see in the next section, this behavior also indicates the
existence of an intermediate dynamical transition between the high
and the low frequency phases.

\subsection{Phase Diagram}
\label{secpd}

In this section, we chart out the different dynamical regimes of the
system. It is well-known that for the closed system ($\lambda=0$),
both the Ising and the Kitaev model exhibits two different dynamical
regimes depending on the drive frequency \cite{dt2,dt3}. For high
drive frequencies, all the non-trivial correlators $C_1(\vec
k,n_0,T) = \langle \psi_k(n_0 T)|c_{\vec k}^{\dagger} c_{\vec
k}|\psi_{\vec k}(n_0T)\rangle$, $C_2(\vec k,n_0,T)= \langle
\psi_k(n_0 T)|c_{-\vec k} c_{-\vec k}^{\dagger}|\psi_{\vec
k}(n_0T)\rangle$, and $C_3(\vec k, n_0,T)= \langle \psi_k(n_0
T)|c_{\vec k}^{\dagger} c_{-\vec k}^{\dagger}|\psi_{\vec
k}(n_0T)\rangle$ decay to their steady state values  (reached as
$n_0 \to \infty$) as $n_0^{-(d+1)/2}$. In contrast, for low drive
frequencies, they decay as $n_0^{-d/2}$. These two regimes are
separated by several reentrant phase transitions at specific
critical frequencies for the Ising model in $d=1$; for the Kitaev
model, there is a single phase transition occurring as the drive
frequency is lowered. The aim of the present section is to study the
fate of these dynamical regime when $\lambda \ne 0$.

To understand why such a transition occur and to decipher its
relation with the structure of the Floquet eigenvalues, we first
rewrite the correlators in terms of the eigenvectors
$|\psi_m^F\rangle = \prod_{\vec k} |\psi_{\vec k m}^F \rangle$ and
eigenvalues $\exp[-i \epsilon_{\vec k m}^F T/\hbar]$ of $U_{\vec
k}(T,0)$ using Eqs.\ \ref{uevol1}. In terms of these, the
wavefunction after $n_0$ drive cycles and for an initial starting
state $|\psi_{\vec k}(0)\rangle$, can be written as
\begin{eqnarray}
|\psi_{\vec k}(n_0)\rangle &=& \sum_{m=1..4} e^{-i n_0 \epsilon_m^F
T/\hbar} \alpha_m |\psi_{m \vec k}^F\rangle \label{wav1}
\end{eqnarray}
where $\alpha_m(\vec k)= \langle \psi_{m \vec k}^F|\psi_{\vec
k}(0)\rangle$ and we have used the representation of $U$ in terms of
its eigenvalues and eigenvectors. Using this one can obtain
\begin{widetext}
\begin{eqnarray}
\delta C_i(\vec k, n_0,T) &=& \sum_{m_1 \ne m_2}
\alpha_{m_1}^{\ast}(\vec k) \alpha_{m_2}(\vec k) \chi_i^{m_1
m_2}(\vec k) e^{in_0(\epsilon^F_{m_1}(\vec k)-\epsilon_{m_2}^F(\vec
k))T/\hbar} = \sum_{m_1,m_2} f_i^{m_1 m_2}(\vec k,n_0,T)
e^{in_0(\epsilon^F_{m_1}(\vec k)-\epsilon_{m_2}^F(\vec
k))T/\hbar}  \nonumber\\
\chi_i^{m_1 m_2} (\vec k) &=&\langle \psi_{m_1 \vec k}^F|{\mathcal
O}_i(\vec k)|\psi_{m_2 \vec k}^F \rangle \label{corrdef1}
\end{eqnarray}
\end{widetext}
where $\delta C_i= C_i - C_i^{{\rm steady \, state}}$ and $i=1,2,3$.
Here ${\mathcal O}_1= c_{\vec k}^{\dagger} c_{\vec k}$, ${\mathcal
O}_2= c_{\vec -k} c_{-\vec k}^{\dagger}$, and ${\mathcal O}_3=
c_{\vec k}^{\dagger} c_{-\vec k}^{\dagger}$. This indicates that in
real space, these correlation functions can be written as
\begin{eqnarray}
\delta C_i(\vec r,n_0,T)= \frac{1}{2} \int \frac{d^d k}{(2 \pi)^d}
e^{ i \vec k \cdot \vec r} \delta C_i(\vec k, n_0,T)
\label{corrdef2}
\end{eqnarray}
To see the behavior of $\delta C_i$ at large $n_0$, we note that for
any function $f_i(k)$ and for large integer $n_0$, one has the
identity
\begin{eqnarray}
&&\int f_i(\vec k) e^{i n_0 \phi(\vec k)} d^d k \approx e^{i n_0
\phi(\vec k_0)}
(n_0 \phi^{''}(\vec k_0))^{\frac{-d}{2}} \nonumber \\
&& \times e^{\frac{\pi i \mu}{4}} \left( f_i(\vec k_0) + i
\frac{f^{''}(\vec k_0)}{2 \phi^{''}(\vec k_0) n_0} + {\rm
O}(1/n_0^2)) \right)
\end{eqnarray}
where $\vec k_0$ is the saddle point such that $\phi'(\vec k_0)=0$.
We find that the leading behavior of this integral will be $\sim
n_0^{-d/2}$ if $f_i(\vec k_0) \ne 0$ and $\sim n_0^{-(d+2)/2}$
otherwise. Using this identity, we find that the behavior of the
correlators $\delta C_i$ comes from the saddle points of the
difference of Floquet eigenvalues which we denote by $\epsilon_{\vec
k}^F$. For saddle points at high frequencies, it may be possible
that $f_i^{m_1,m_2}(\vec k_0^F,n_0,T)=0$ for all $m_1$ and $m_2$.
This typically happens when these saddles occur at the center or
edge of the Floquet Brillouin zone and leads to a $1/n_0^{(d+2)/2}$
decay of the correlators. At lower frequencies, the contribution of
the correlators comes from the saddles which are not necessarily at
the zone edge or center and these lead to $1/n_0^{d/2}$ decay of the
correlators. The transition between these two phases occur at the
critical frequency where an extrema first occurs in the Floquet
spectrum away from the zone edge or center. This transition was
shown to exist for the closed system for both Ising and Kitaev model
in Ref.\ \onlinecite{dt2}. Here we are going to numerically
investigate its fate in the presence of a fermionic bath.

\begin{figure}[H]
\centering
\includegraphics[width=0.48\linewidth ,height=0.39\columnwidth]{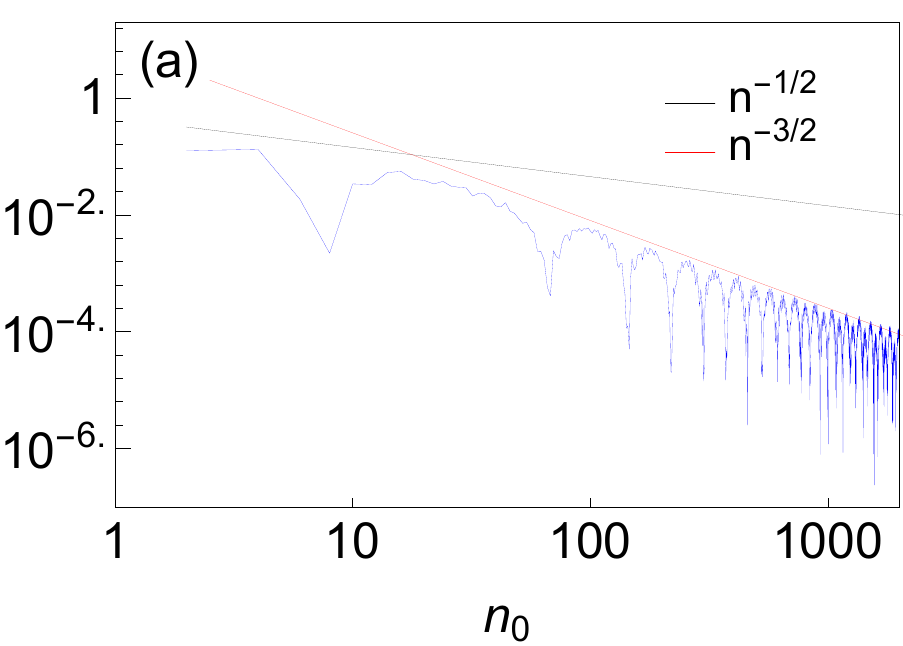}
\includegraphics[width=0.48\linewidth ,height=0.39\columnwidth]{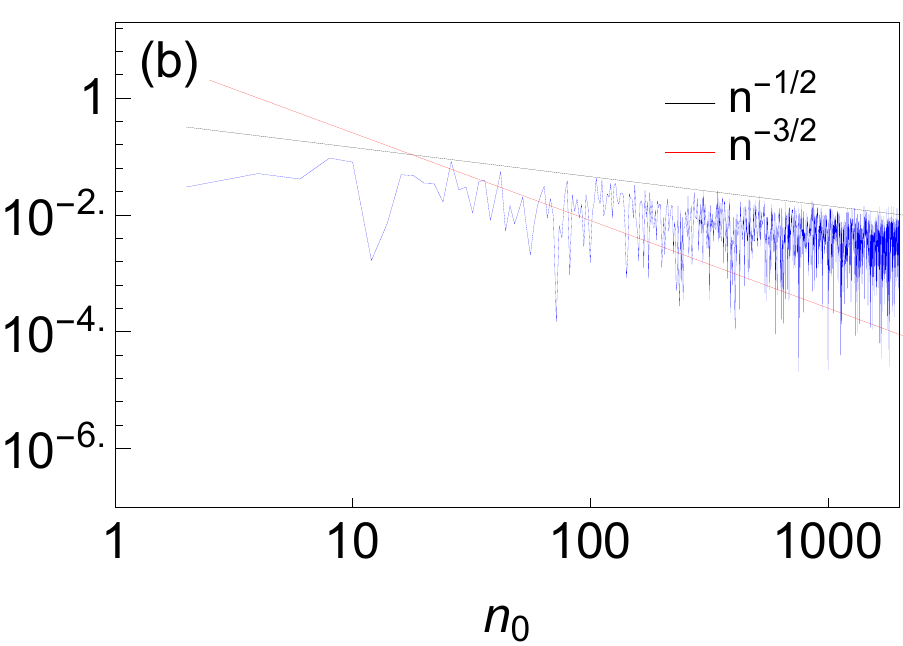}
\caption{Plot of the correlator $\delta C_1(\vec r=0,n_0,T)$ as a
function of $n_0$ for (a) $ \omega_D=10 \pi$  and (b) $
\omega_D = 0.2 \pi$ for the 1D Ising model. For both plots, $g_f=2$,
$g_i=0$, $\lambda=0.8$, and $\eta=0.1$. All energies(frequencies)
are in units of $J(J/\hbar)$. See text for details.} \label{fig4}
\end{figure}

To this end, we first consider the 1D Ising model, where we show the
evolution of $\delta C_1(\vec r=0, n_0,T)$ as a function of $n_0$ in
Fig.\ \ref{fig4} for (a)$\hbar \omega_D/J=10 \pi$  and
(b) $\hbar \omega_D/J=0.2 \pi$. For the plot we have chosen
$g_f/J=2$, $g_i=0$, $\lambda/J=0.8$ and $\eta/J=0.1$. Fig.\
\ref{fig4} clearly demonstrate two dynamical regimes; in the high
frequency regime, the correlators decay to their steady state value
as $n_0^{-3/2}$ while for the low-frequency phase, they have a
$n_0^{-1/2}$ behavior. This shows that the different dynamical
regimes persists in the presence of fermionic bath for the 1D Ising
system. We have checked that similar behavior is  seen for both
$\delta C_{2}$ and $\delta C_3$.

\begin{figure}[H]
\centering
\includegraphics[width=0.45\linewidth ,height=0.41\columnwidth]{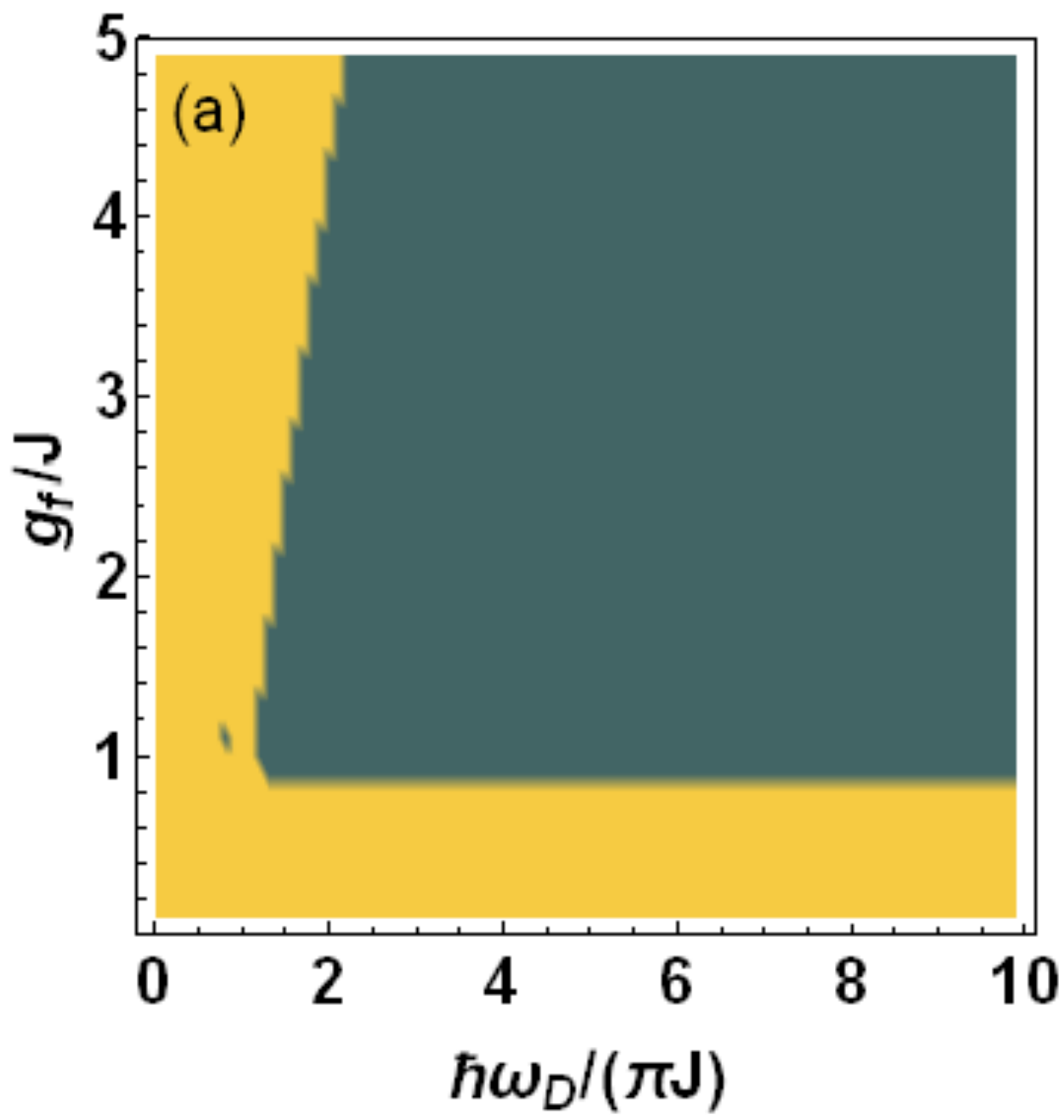}
\includegraphics[width=0.45\linewidth ,height=0.41\columnwidth]{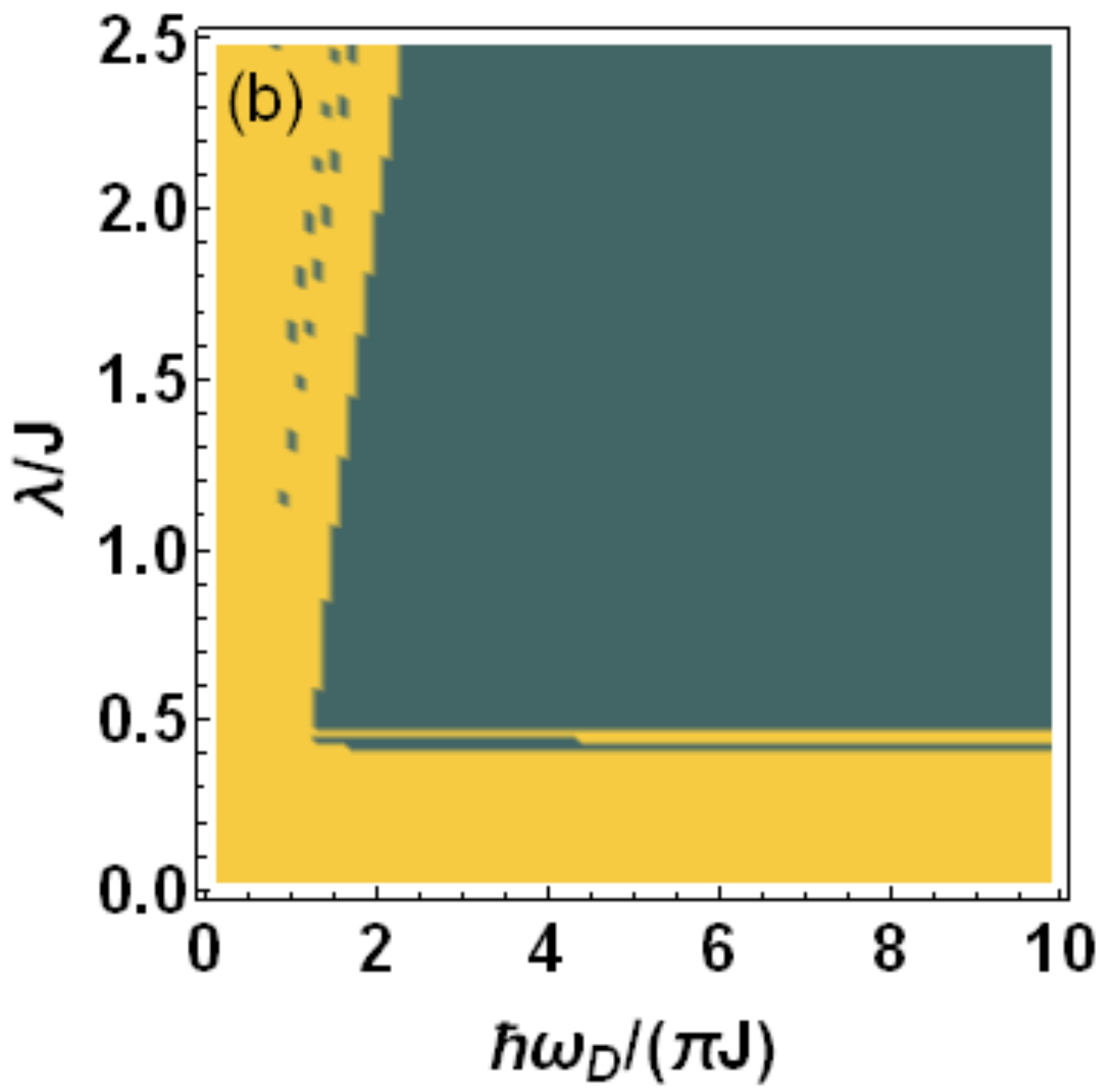}
\includegraphics[width=0.49\linewidth ,height=0.41\columnwidth]{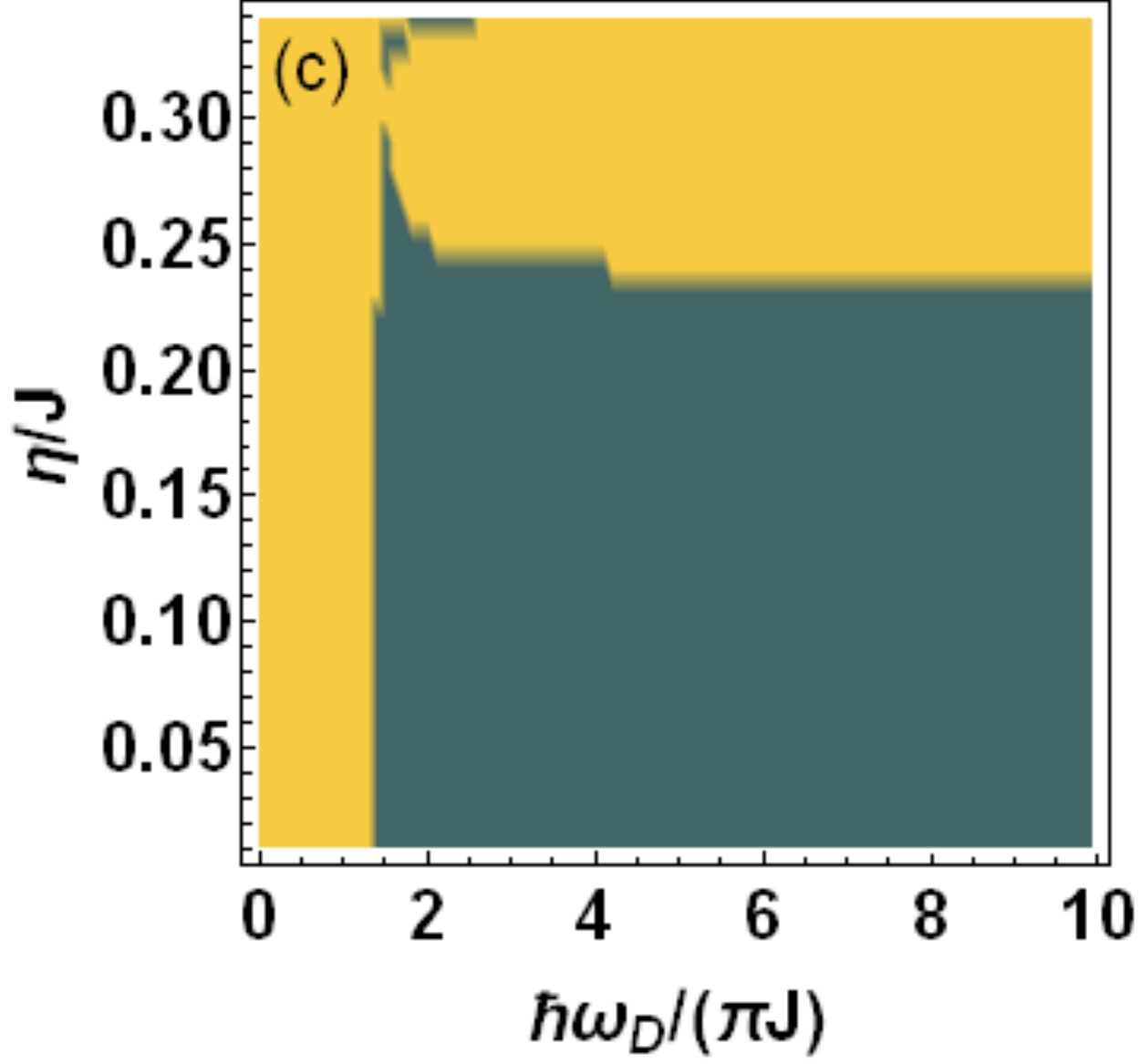}
\includegraphics[width=0.47\linewidth ,height=0.41\columnwidth]{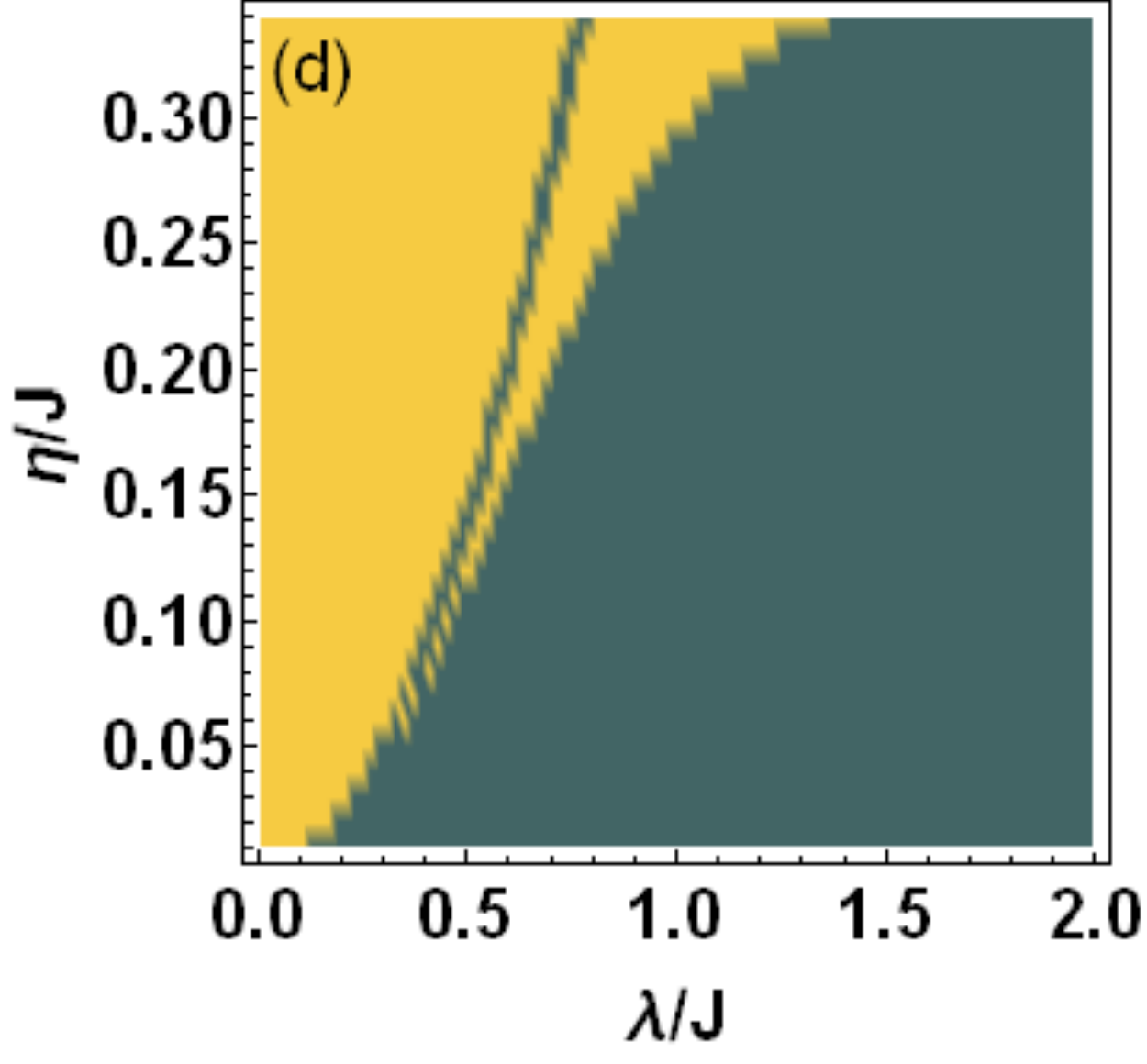}
\caption{(a) Plot of different dynamical regimes as a function
of $g_f/J$ and $\hbar \omega_D/(\pi J)$ for $\lambda=0.8J$ and
$\eta=0.1 J$. The green [yellow] region corresponds to
$n_0^{-3/2}\,[n_0^{-1/2}]$ behavior of the correlators.(b) Plot
of the dynamical regimes as a function of $\lambda/J$ and $\hbar
\omega_D/(\pi J)$ for $\eta=0.1 J$ and $g_f=2J$.
(c) Plot of the dynamical regimes as a function of $\eta/J$ and $\hbar
\omega_D/(\pi J)$ for $\lambda=0.8J$ and $g_f=2J$. (d)
Plot of the dynamical regimes as a function of $\eta/J$ and
$\lambda/J$ for $\hbar \omega_D/(\pi J)=10$ and $g_f=2J$. For all
plots $g_i=0$ and $J=1$. See text for details.} \label{fig5}
\end{figure}

\begin{figure}[H]
\centering
\includegraphics[width=0.48\linewidth ,height=0.35\columnwidth]{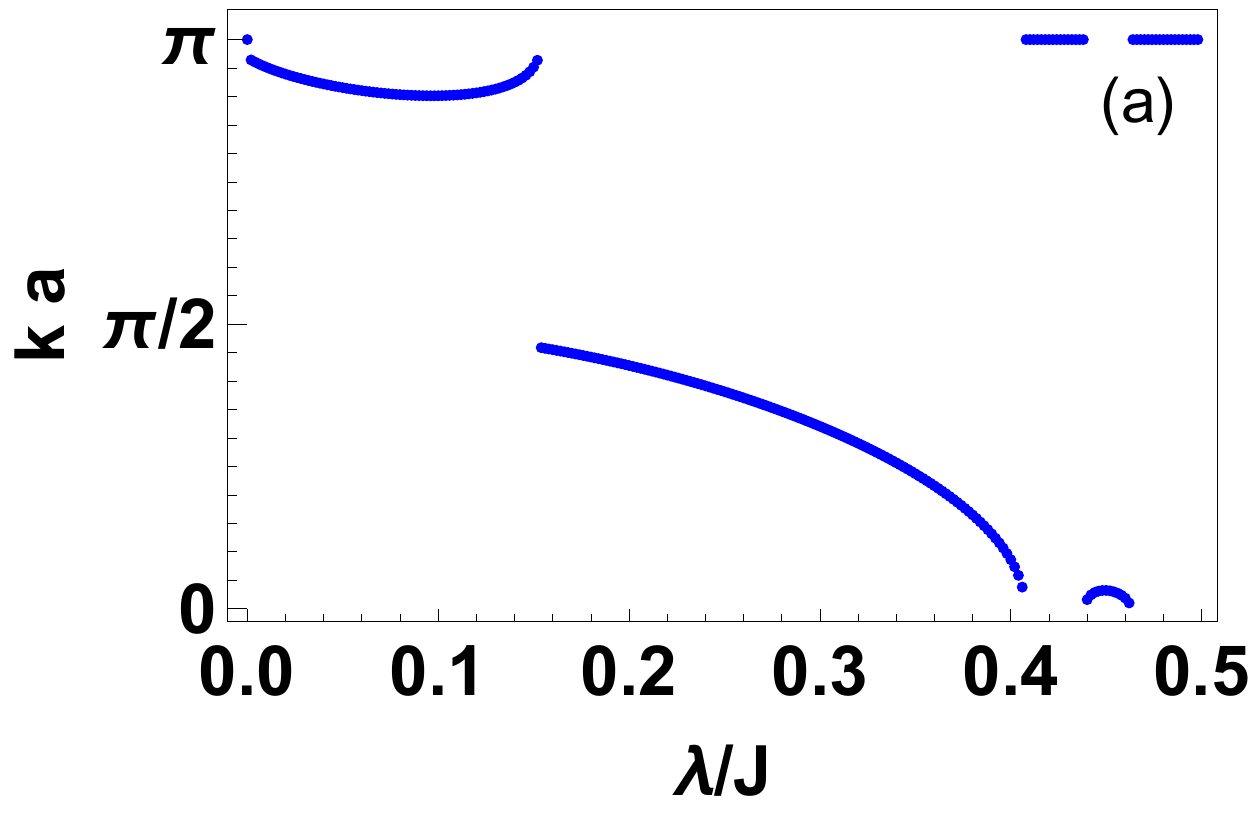}
\includegraphics[width=0.48\linewidth ,height=0.35\columnwidth]{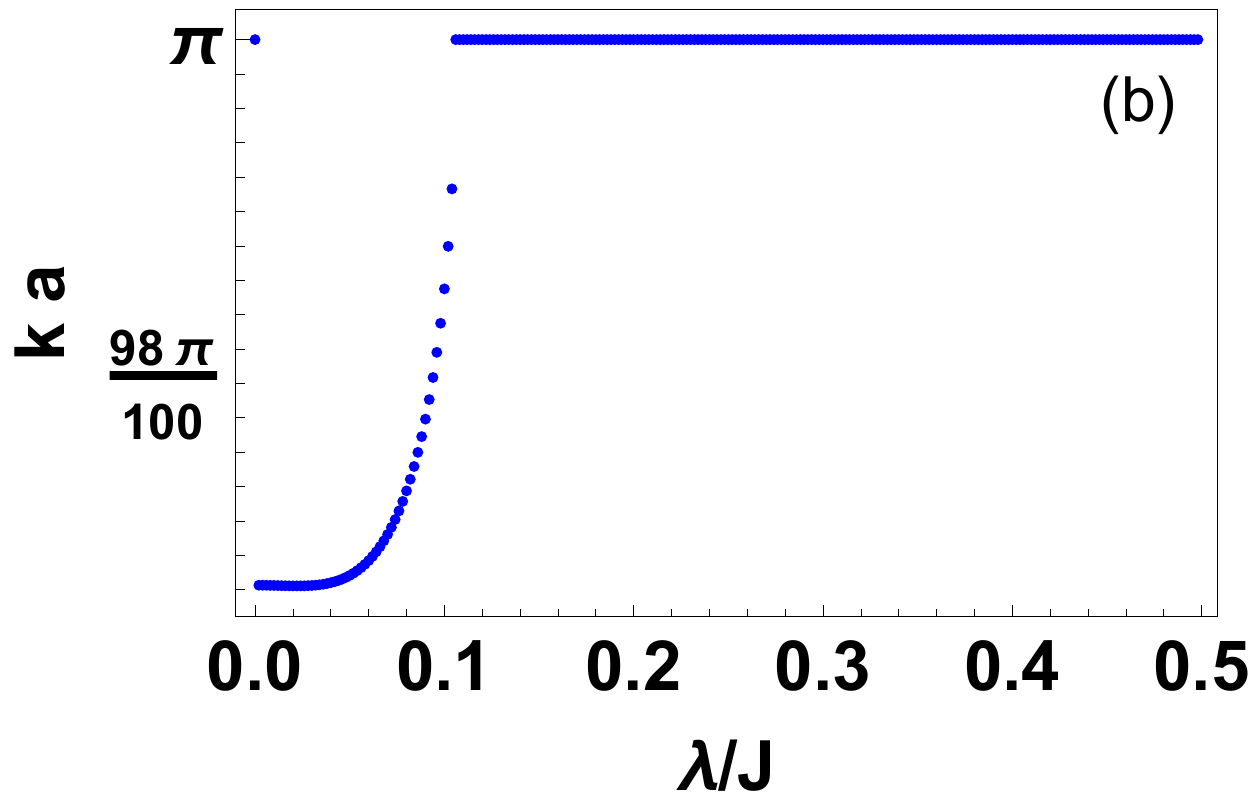}
\caption{The position of extrema of two of the Floquet eigenvalues
as a function of $\lambda/J$ for $\eta=0.1 J$, $g_f=2J$, $g_i=0$,
and $\hbar \omega_D/J=10 \pi$. The other two eigenvalues always show
extrema at $k=\pi/a$. In the plot, in case of multiple extrema, the
one with the highest momenta other than $\pi/a$ is shown. For all
plots, $J=a=1$.} \label{fig6}
\end{figure}

The corresponding phase diagram displaying the two different
dynamical regimes is shown in Fig.\ \ref{fig5}. Fig.\ \ref{fig5}(a)
shows these regimes as a function of $g_f$ and $\omega_D$ for
$\eta=0.1J$ and $\lambda=0.8 J$. The plot demonstrates the presence
of dynamical transition for finite $\lambda$. We note that in the
presence of finite $\lambda$ and $\eta$, one needs a finite critical
drive amplitude $g_{f}= g_{fc} \simeq 0.9 J$ for the transition to
occur; for $g_f < g_{fc}$ only the low frequency regime with $\delta
C_i \sim n_0^{-1/2}$ survives. Furthermore, for a small window of
$1.1\le g_f \le 1.2$, we find the presence of reentrant transitions
as a function of $\omega_D$ with the second transition taking place
around $\hbar \omega_D \sim 0.9 J$. Fig.\ \ref{fig5}(b) shows the
position of these dynamical regimes in the $\lambda -\omega_D$ plane
for a fixed $\eta=0.1 J$, $g_f=2J$ and $g_i=0$. We note that the
presence of a small $\lambda$ leads to $n_0^{-1/2}$ decay of the
correlators even when the closed system at $\lambda=0$ exhibits
$n_0^{-3/2}$ behavior. This can be further understood by noting the
behavior of the Floquet eigenvalues for the Ising model; the extrema
of these eigenvalues shifts from $\pi$ for infinitesimal $\lambda$
as shown in Fig.\ \ref{fig6}. Up on increasing
$\lambda$, the extrema returns to $\pi$ for $\lambda/J \simeq 0.45$
for $\hbar\omega_D/J \ge 1.8 \pi$ as can be seen from Fig.\
\ref{fig6}; this leads to the presence of dynamical transition at
large enough $\lambda$ even when the small $\lambda$ regime has no
such transition. Moreover, one finds that at high frequencies $\hbar
\omega_D/J \ge 1.8 \pi$, it is possible to have multiple transitions
between the two dynamical regimes by tuning the coupling to the bath
at a fixed frequency; this phenomenon has no analog in closed system
studied earlier.

Fig.\ \ref{fig5}(c) demonstrate the
dependence of these regimes on $\eta$ and $\omega_D$ for $g_f=2J$,
$g_i=0$, and $\lambda=0.8 J$. We find that there is a wide range of
$\eta$ for which the transition is stable.

Finally we chart out the position of these dynamical regimes in the
$\eta-\lambda$ plane for a fixed drive frequency $\hbar \omega_D =
10 \pi J$, $g_f=2J$, and $g_i=0$ in Fig.\ \ref{fig5}(d). We find the presence of both
dynamical regimes as a function of $\eta$ and $\lambda$ and multiple
transition curves separating them. We note that only the regime with
$n_0^{-1/2}$ behavior persists for small $\lambda$ which is
consistent with the behavior of the Floquet eigenvalues in Fig.\
\ref{fig6}.

\begin{figure}[H]
\centering
\includegraphics[width=0.48\linewidth ,height=0.39\columnwidth]{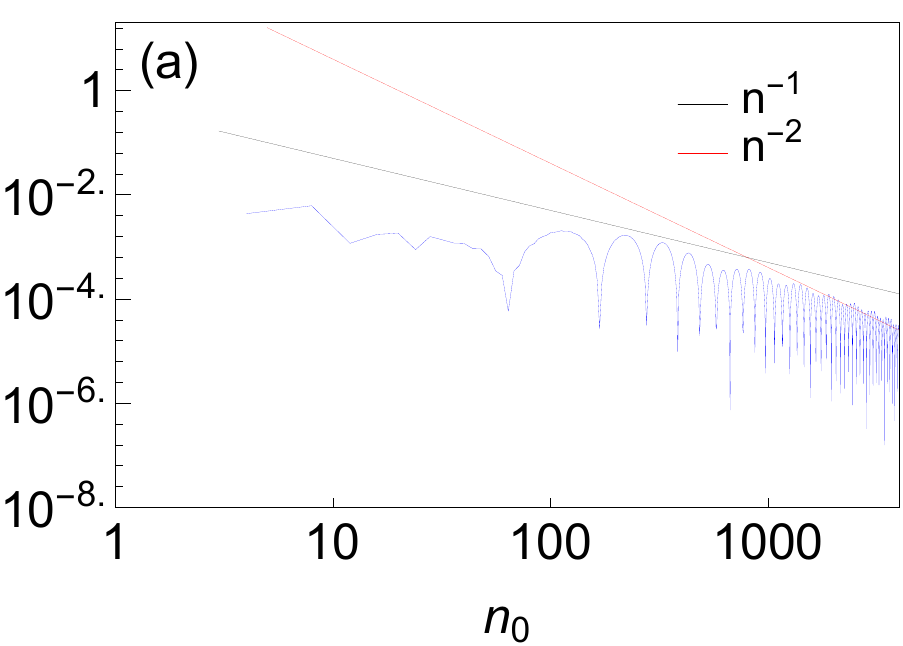}
\includegraphics[width=0.48\linewidth ,height=0.39\columnwidth]{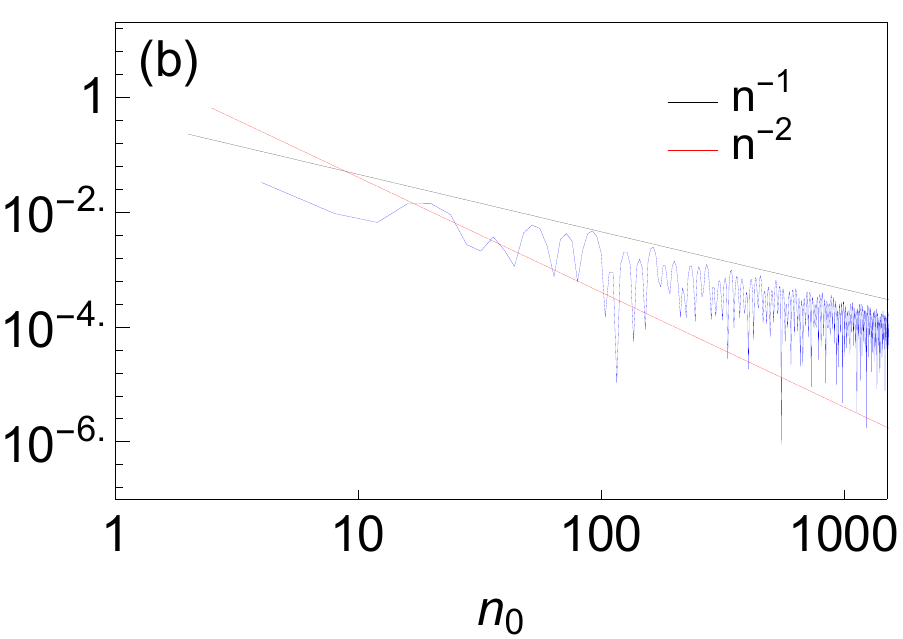}
\caption{Plot of $\delta C_1$ as a function of $n_0$ for $\eta=0.1
J_1$, $J_{3f}=5J_1$, $J_{3i}=4J_1$ $\lambda=0.8 J_1$ and $\hbar
\omega_D/(\pi J_1)=10[3.3]$ for (a)[(b)]. For all
plots, $J_1=J_2=1$. See text for details} \label{fig7}
\end{figure}

Next, we chart out the phase diagram for the Kitaev model. To this
end, in Fig.\ \ref{fig7}, we plot $\delta C_1(n_0,T)$ as a function
of $n_0$ for $\hbar \omega_D/J_1 =10 \pi$ (Fig.\ \ref{fig7}(a)) and $\hbar
\omega_D/J_1=3.3 \pi$ (Fig.\ \ref{fig7}(b)), $\lambda/J_1=0.8$, $J_2/J_1=1$,
$J_{3 f[i]}/J_1 = 4[5]$, and $\eta/J_1=0.1$. We find that Fig.\ \ref{fig7}(a)
shows a $1/n_0^2$ decay while Fig.\ \ref{fig7}(b) exhibits $1/n_0$
behavior; this constitutes a clear signature of dynamical transition
for finite $\lambda$ and $\eta$. We have checked that the behavior
of $\delta C_2$ and $\delta C_3$ are similar to $\delta C_1$.

\begin{figure}[H]
\centering
\includegraphics[width=0.44\linewidth ,height=0.41\columnwidth]{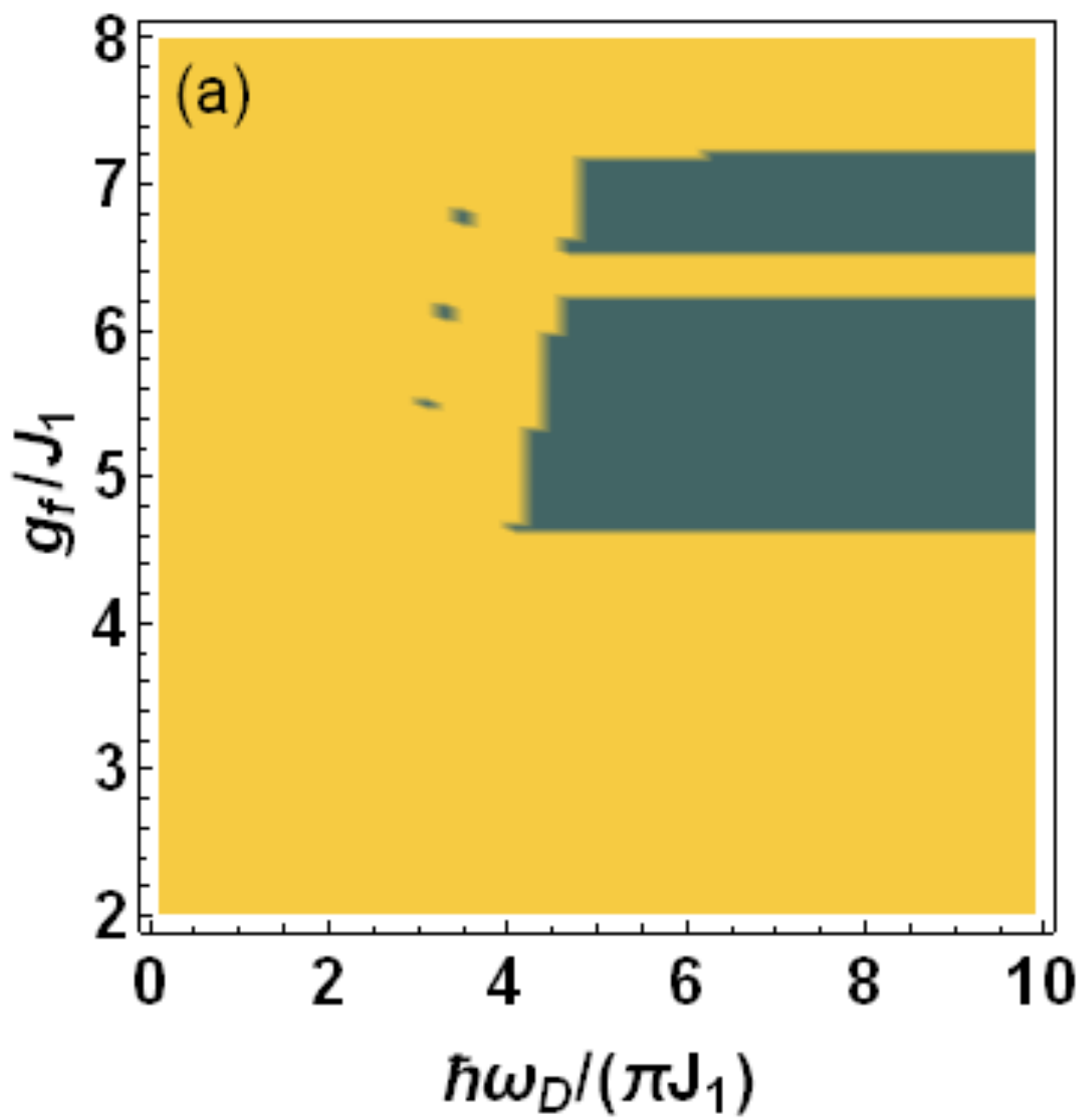}
\includegraphics[width=0.45\linewidth ,height=0.41\columnwidth]{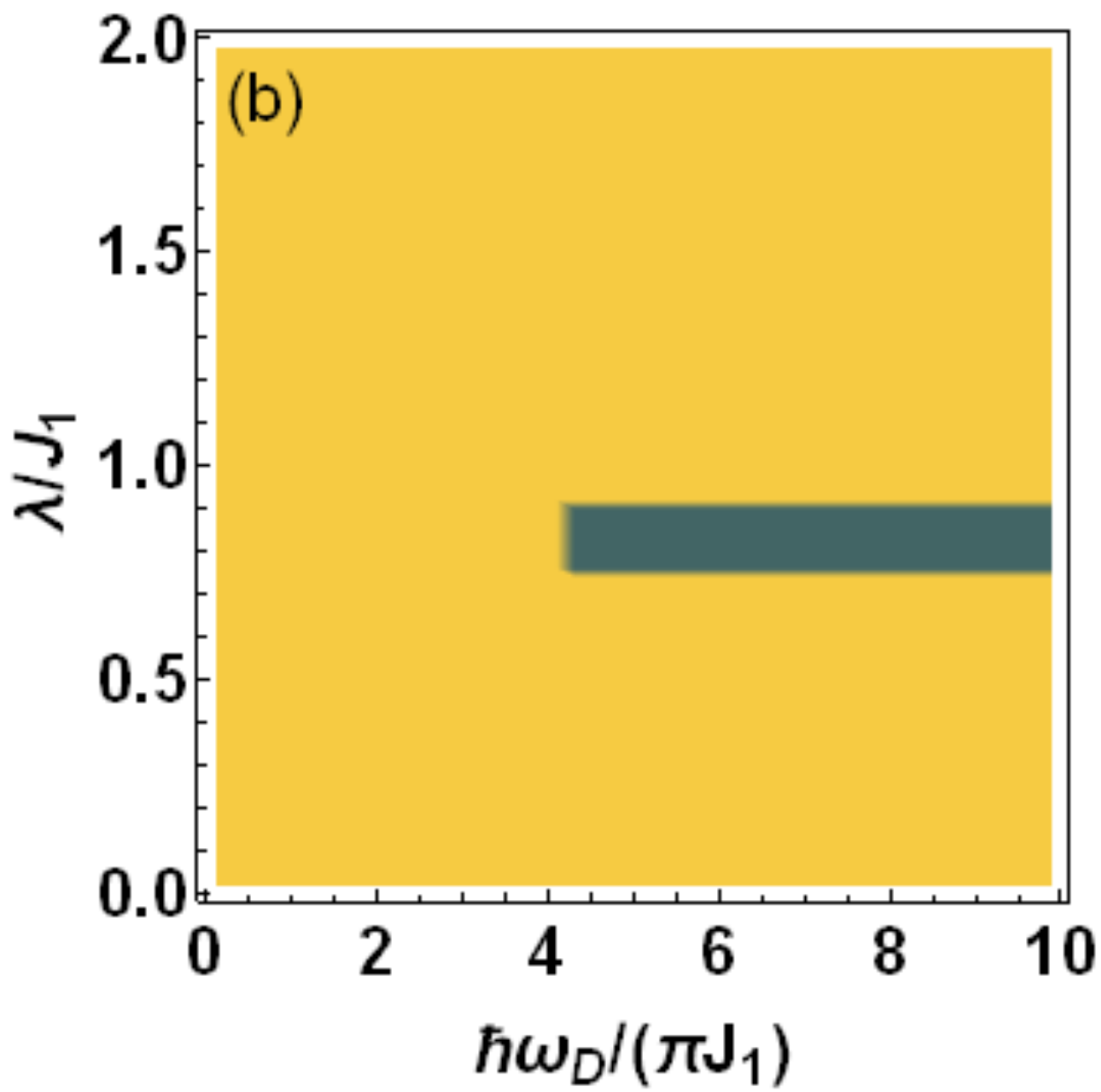}
\includegraphics[width=0.49\linewidth ,height=0.41\columnwidth]{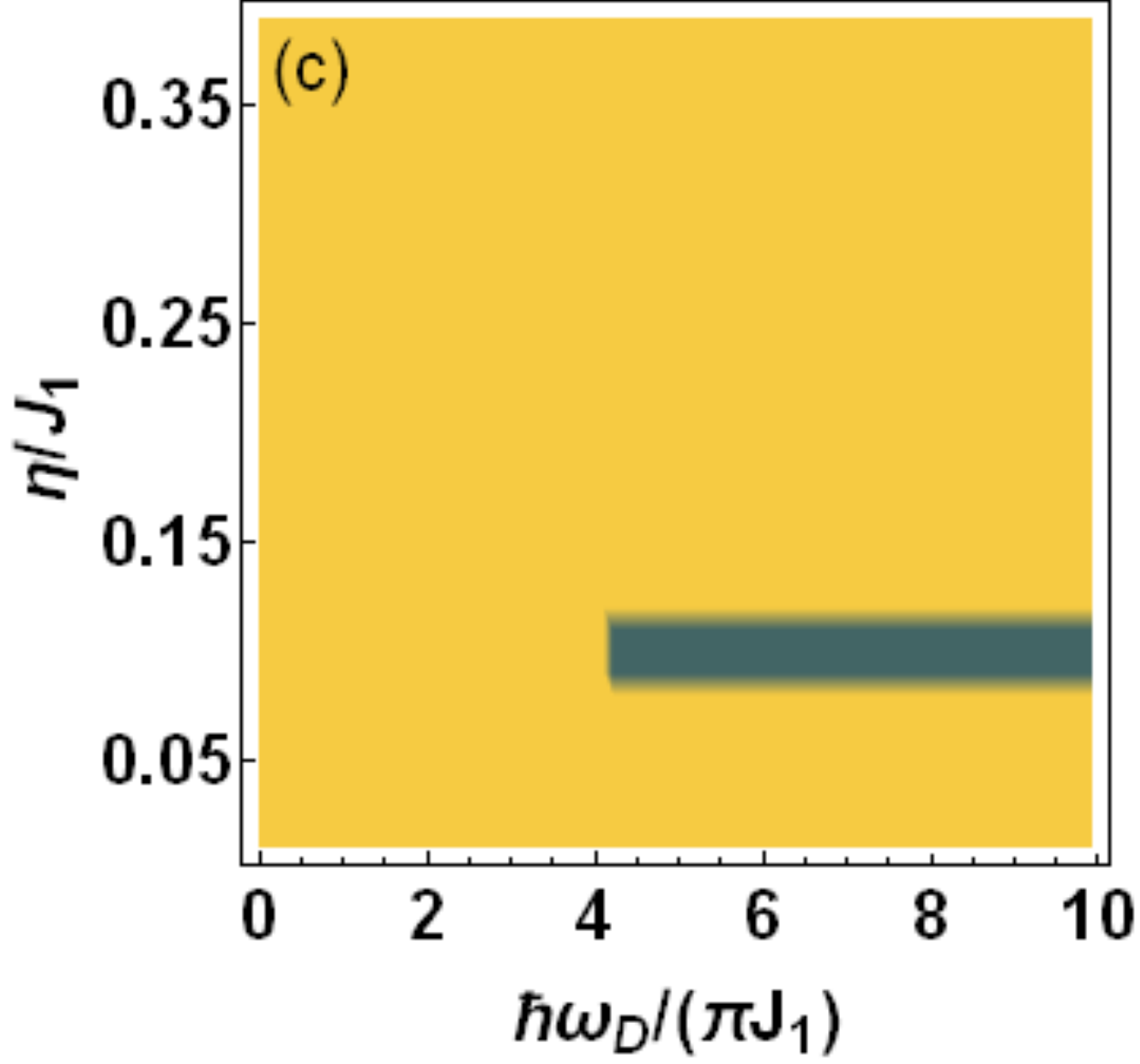}
\includegraphics[width=0.47\linewidth ,height=0.41\columnwidth]{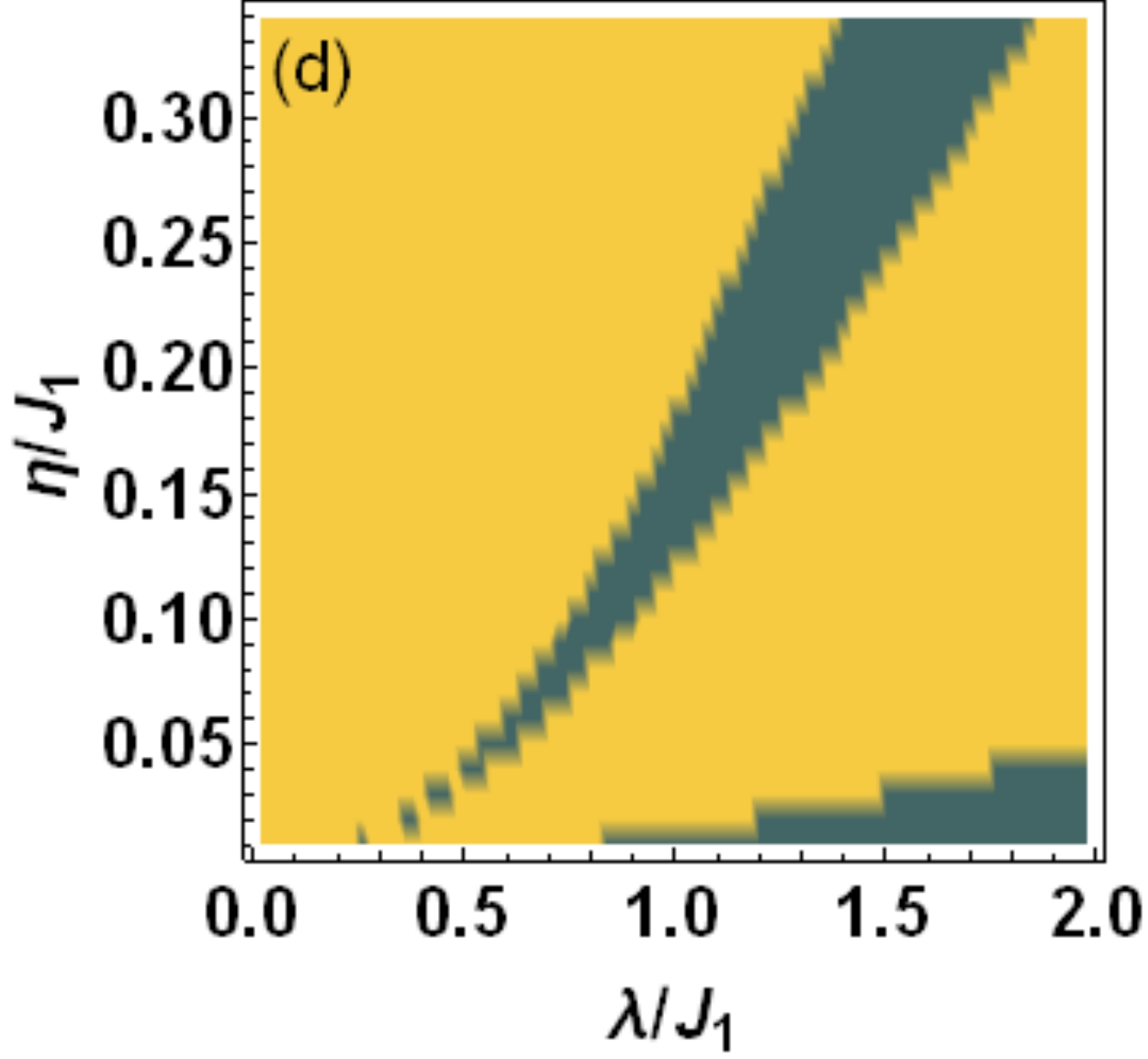}
\caption{(a) Plot of different dynamical regimes as a function
of $J_{3f}/J_1 \equiv g_f$ and $\hbar \omega_D/(\pi J_1)$ for
$\lambda=0.8J_1$ and $\eta=0.1 J_1$. The green [yellow] region
corresponds to $n_0^{-2}\,[n_0^{-1}]$ behavior of the correlators.
(b) Plot of the dynamical regimes as a function of $\eta/J_1$
and $\hbar \omega_D/(\pi J_1)$ for $\lambda=0.8J_1$ and $J_{3f}=5
J_1$. (c) Plot of the dynamical regimes as a function of
$\lambda/J_1$ and $\hbar \omega_D/(\pi J_1)$ for $\eta=0.1 J_1$ and
$J_{3f}=5J_1$. (d) Plot of the dynamical regimes as
a function of $\eta/J_1$ and $\lambda/J_1$ for $\hbar \omega_D/(\pi
J_1)=10$ and $J_{3f}=5J_1$. For all plots $J_{3i}=4J_1$ and
$J_2=J_1$. See text for details.} \label{fig8}
\end{figure}

The phase diagram displaying different dynamical regimes is
exhibited in Fig.\ \ref{fig8}. In all of these plots we choose
$J_2=J_1$ and $J_{3i}/J_1=4$. Fig.\ \ref{fig8}(a) shows the
dynamical regimes as a function of $J_{3f}/J_1\equiv g_f$ and $\hbar
\omega_D/(\pi J_1)$ for $\eta/J_1=0.1$ and $\lambda/J_1=0.8$. We
find that for a distinct range of $J_{3f}$, the system displays both
$1/n_0^2$ (green regions) and $1/n_0$ (yellow regions) behavior.
This constitutes examples of dynamical transition. In Fig.\
\ref{fig8}(b), we chart out these dynamical regimes as a function of
$\lambda/J_1$ and $\hbar \omega_D/(\pi J_1)$ for $J_{3f}/J_1=5$ and
$\eta/J_1=0.1$. We find that the 1/$n_0^2$ behavior can only be seen
within a finite range $ 0.75 \le \lambda/J_1 \le 0.92$. The extent
of this region depends on $\eta/J_1$; it becomes wider with larger
$\eta$. In Fig.\ \ref{fig8}(c), we plot the dynamical regimes as a
function of $\eta/J_1$ and $\hbar \omega_D/(\pi J_1)$ for
$\lambda/J_1=0.8$ and $J_{3f}/J_1=5$. We find that for
$\lambda/J_1=0.8$ there is a narrow region in $\eta$ where $1/n_0^2$
behavior survives. Finally Fig.\ \ref{fig8}(d), we plot the position
of these dynamical regimes in the $\eta-\lambda$ plane for $\hbar
\omega_D/(\pi J_1)=10$ and $J_{3f}/J_1=5$. We find that increasing
$\eta$ shifts the presence of the dynamical regime with $1/n_0^2$
behavior to higher values of $\lambda$; it also makes it extent
wider. Furthermore, for small $\eta/J_1$, a larger $\lambda/J_1 >0.8
$ allows presence of $1/n_0^2$ decay of correlators. Also, we find
that increasing $\eta$ with $\lambda/J_1>0.8$ leads to reentrant
transition between the two dynamical regimes; these transitions do
not have any analogue in closed systems studied earlier in Ref.\
\onlinecite{dt2,dt3}.

\section{Bosonic Bath}
\label{boson}

In this section we couple $H_0$ to a bosonic bath. The technique
used for obtaining our result is detailed in Sec.\ \ref{eomsec}
while the numerical results are presented in Sec.\ \ref{numres1}.

\subsection{Equation of motion}
\label{eomsec}

In the presence of a bosonic bath, the total Hamiltionian of the
system  reads
\begin{eqnarray}
H_{\rm total}=& H_{0}(t) +H'_b+H'_{\rm int} \label{htot2}
\end{eqnarray}
where the bath Hamiltonian, modeled by a bunch of harmonic
oscillators, is given by
\begin{eqnarray}
H'_b &=& \sum_{\vec q} \hbar \omega_{\vec q}  b_{\vec q}^{\dagger}
b_{\vec q}. \label{bbathham}
\end{eqnarray}
Here $b_{\vec q}^{\dagger}$ is creation operator for bosons and
$\omega_{\vec q}$ is the corresponding frequency. The interaction
between the Fermions and the bath is given by
\begin{eqnarray}
H'_{\rm int}&=& \sum_{\vec k \vec q}  \lambda_{\vec k} c_{\vec
k}^{\dagger} c_{\vec k+ \vec q} (b_{\vec q}^{\dagger} + b_{-\vec q})
+{\rm h.c.} \label{bintham}
\end{eqnarray}
where $\lambda_{\vec k}$ is the coupling function which determines
the strength of interaction between the system fermions and the bath
bosons. Here, and in rest of this section, we shall extend
definitions of $c_{\vec k}$ and $c_{-\vec k}$ over the entire
Brillouin zone for convenience; the double counting which arises due
to such an extension can be simply offset by a factor of $1/2$ while
evaluating sum over momentum for computing any correlation
functions.

We note that the interaction between the bosonic bath and the
fermions (Eq.\ \ref{bintham}) necessarily destroys integrability of
the fermion system upon integrating the bath degrees of freedom.
This is in contrast to the case of fermionic bath studied earlier
and, as we shall see, leads to qualitative difference in the
dynamics of correlators of the driven model. In what follows, we
shall study the dynamics of $H_{\rm total}$ ignoring backreaction of
the system to the bath \cite{brref1}. This approximation has been
widely used in treating such open quantum systems; it produces
accurate results for system dynamics when the bath is either very
large compared to the system size or if the bath frequencies are
much larger compared to the system energy scales. In what follows we
shall restrict ourselves to the latter case ($\hbar \omega_{\vec q}$
being the largest energy scale) and assume a thermal distribution
for the bath bosons with a fixed temperature $T_b$ at all times:
$n^b[\omega_{\vec q}] = (\exp[\hbar \omega_{\vec q}/(k_B
T_b)]-1)^{-1}$. We also note that in this limit it is possible to
integrate out the bath degrees of freedom and obtain an effective
static interaction between the fermions with strength $\sim
\lambda^2/(\hbar \omega_{\vec q})$; thus our analysis also yields
information about dynamics of interacting driven fermions.

To study the dynamics, we note that since the fermionic system, in
the presence of the bath does not reduce to Gaussian action, the
path integral procedure of the previous section can not be applied
here in a straightforward manner and we need to resort to some
approximation scheme. To this end, we use the equation of motion
approach where one writes down the equation of motion for the
correlation functions of the fermions. This, of course leads to an
infinite hierarchy of equations which needs to be truncated. Several
such truncation schemes are discussed in the literature in various
contexts \cite{book1,book2,damp,knorr}. Here we truncate these
equations by writing all four point correlations (for both fermion
and mixed correlators) as a product of two point correlations by
using Wick's theorem.

The Heisenberg equations for expectation of any operator ${\mathcal
O}_{\vec k}$ can then be written as
\begin{eqnarray}
i \partial_t \langle O_{\vec k} \rangle = \langle [H_{\rm total},
{\mathcal O}_{\vec k} ] \rangle \label{eom1}
\end{eqnarray}
where the expectation is taken with respect to the initial state at
$t=0$. Here we shall choose this state to be a direct product state
of fermions and the bath bosons $|\psi\rangle_{\rm init} =
|\psi\rangle_{\rm fermion} \otimes |\psi\rangle_{\rm bath}$. This
procedure yields
\begin{widetext}
\begin{eqnarray}
i \partial_t n(\vec{k}) &=& - \Delta_{\vec k} F^{\ast}(\vec{k})+
\Delta_{\vec k}^{\ast} F(\vec{k}) + \lambda_{\vec k} (
A_1(\vec{k},\vec{q}) +
A_2(\vec{k},\vec{q}) )- \lambda_{\vec{k}-\vec{q}} (A_1(\vec{k}- \vec{q},\vec{q}) + A_2(\vec{k}- \vec{q},\vec{q})) \nonumber \\
 &-& \lambda_{\vec k} ( A_1^{\ast}(\vec{k},\vec{q}) +
A_2^{\ast}(\vec{k},\vec{q}) )
+ \lambda_{\vec{k}- \vec{q} } (A_1^{\ast}(\vec{k}- \vec{q},\vec{q}) + A_2^{\ast}(\vec{k}- \vec{q},\vec{q}) \nonumber \\
i \partial_t F(\vec k) &=& 2 (g(t)- z_{\vec k}) F(\vec k) +
\Delta_{\vec k}( ( n(\vec k) + n(-\vec k) ) -   1) + \lambda_{\vec
k}
( G_1( \vec k,\vec{q}) + G_2 (\vec k,\vec q ))  \nonumber \\
&+& \lambda_{-\vec{k}} (G_1 (\vec k-\vec q,\vec q) + G_2 (\vec
k-\vec q,\vec q) ) + \lambda_{\vec k- \vec q} ( G_1( \vec
k,-\vec{q}) + G_2 (\vec k,-\vec q ))
+ \lambda_{\vec k - \vec q} (G_1 (\vec k+\vec q,-\vec q) + G_2 (\vec k+\vec q,-\vec q) ) \nonumber \\
i \partial_t A_{1[2]} (\vec k, \vec q) &=&  (z_{\vec k} - z_{\vec k+ \vec
q} ) A_{1[2]}( \vec k,\vec q) +   \Delta_{\vec k+
\vec q} G_{2[1]}^{\ast} ( -\vec k, -\vec q)+ \Delta_{\vec k}^{\ast}
G_{1[2]} (\vec k, \vec q) -[+] \hbar \omega_{\vec q} A_{1[2]} (\vec k, \vec q)\nonumber\\
&+& \lambda_{\vec k} (0[1] + n_{b}[+[-]\omega_{\vec q}])(n(\vec k) -
n(\vec k+ \vec q)) -[+]\lambda_{\vec k} n(\vec k+ \vec q) +[-]
\lambda_{\vec k} n(\vec k) n(\vec k+ \vec q) -[+]
\lambda_{- \vec k - \vec q} F^{\ast}(\vec k) F(\vec k+ \vec q) \nonumber \\
&& - i \gamma_0  A_{1[2]}( \vec k,\vec q) \nonumber \\
i \partial_t G_{1[2]}( \vec k, \vec q)&=&( 2 g(t)- z_{\vec k+ \vec q} -
z_{\vec k}  ) G_{1[2]}( \vec k, \vec q) +  \Delta_{\vec k}
A_{1[2]} (\vec k, \vec q) +  \Delta_{\vec k+ \vec q} A_{1[2]}(-\vec k - \vec q, \vec q)
-[+] \omega_{\vec q} G_{1[2]}( \vec k, \vec q) \nonumber \\
&+&  (0[1] + n_b[+[-]\omega_{\vec q}]) ( \lambda_{\vec k} F(\vec k)
+ \lambda_{-\vec k- \vec q} F(\vec k+ \vec q)) +[-] \lambda_{- \vec
k - \vec q} n(-\vec k)F(\vec k+ \vec q) +[-] \lambda_{\vec k} F(\vec
k) n(\vec k+\vec q)\nonumber \\ &-& i \gamma_0  G_{1[2]}( \vec k,
\vec q) \label{bosoneq1}
\end{eqnarray}
\end{widetext}
where the terms $\sim \gamma_0$ has been added to the equations of
the correlators to counter numerical instability arising from
truncation of the hierarchy as discussed earlier \cite{book1,damp}.
The correlators in Eq.\ \ref{bosoneq1} are given by
\begin{eqnarray}
n(\vec k) &=& \langle c_{\vec k}^{\dagger} c_{\vec k} \rangle, \quad
F(\vec k) = \langle c_{-\vec k}  c_{\vec k} \rangle
\label{opdef1} \\
A_1( \vec k,\vec q) &=& \langle c_{\vec k}^{\dagger} c_{\vec k+ \vec
q} b_{\vec q}^{\dagger}\rangle ,\quad  A_2( \vec k, \vec q)
=\langle c^{\dagger}_{\vec k -\vec q} c_{\vec k} b_{-\vec q} \rangle \nonumber\\
G_1( \vec k,\vec q) &=& \langle c_{-\vec k} c_{\vec k+ \vec q}
b_{\vec q}^{\dagger} \rangle, \quad G_2( \vec k, \vec q) = \langle
c_{-\vec k} c_{\vec k+ \vec q} b_{-\vec q} \rangle \nonumber
\end{eqnarray}
We note from Eq.\ \ref{bosoneq1} that the equations for the two
point correlators such as $n_{\vec k}$ and $F_{\vec k}$ gives rise
to higher order mixed correlators $A_{1,2}( \vec k,\vec q)$ and
$G_{1,2} ( \vec k,\vec q)$ which
quantify correlations between electrons and phonons. These mixed
correlators, in turn, give rise to four fermion terms which have
been decomposed into lower order two point correlators using Wick's
theorem as mentioned earlier. This leads to the closed set of
equations (Eq.\ \ref{bosoneq1}) which are solved numerically to
study the dynamics.

\subsection{Numerical Results}
\label{numres1}

The numerical solution of Eq.\ \ref{bosoneq1} allows us to obtain
information about dynamics of both Ising and Kitaev models coupled to
bosonic bath. For all numerical solutions used for results presented
in this section, we have set $\lambda_{\vec k} = \lambda$ for all
$\vec k$ and, unless otherwise mentioned, kept the phenomenological
damping constant $\gamma_0 =0.2 \lambda$, where $\lambda/J$ is
considered to be the smallest scale in the problem. We have checked,
by varying $\gamma_0$ around this value, that the nature of the
correlator remains independent of $\gamma_0$ value in this regime.
Also, for all plots, we have used a single bosonic mode at $\vec
q=\vec q_0=(4\pi/L,4\pi/L)$ (where $L$ is the linear dimension of
the system) for the Kitaev model, $q=q_0 =4 \pi/L$ for the Ising
model, and have set $\hbar \omega_{q_0}/J=20$ to be the largest
scale in the problem. We have chosen a finite non-zero $\vec q_0$ to
ensure non-trivial coupling to the bath (for $\vec q=0$, $[n_{\vec
k},H_1]=0$) while $\hbar \omega_{\vec q}/J \gg1$ is chosen to ensure
that neglecting back-reaction of the system on the bath remains a
valid assumption.

\begin{figure}[H]
\centering
\includegraphics[width=0.48\linewidth ,height=0.38\columnwidth]{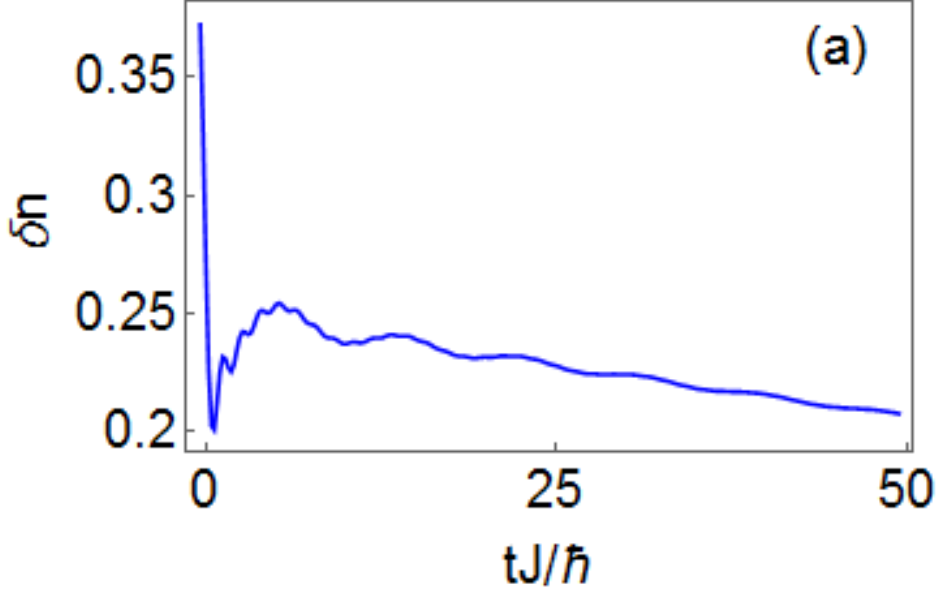}
\includegraphics[width=0.48\linewidth ,height=0.38\columnwidth]{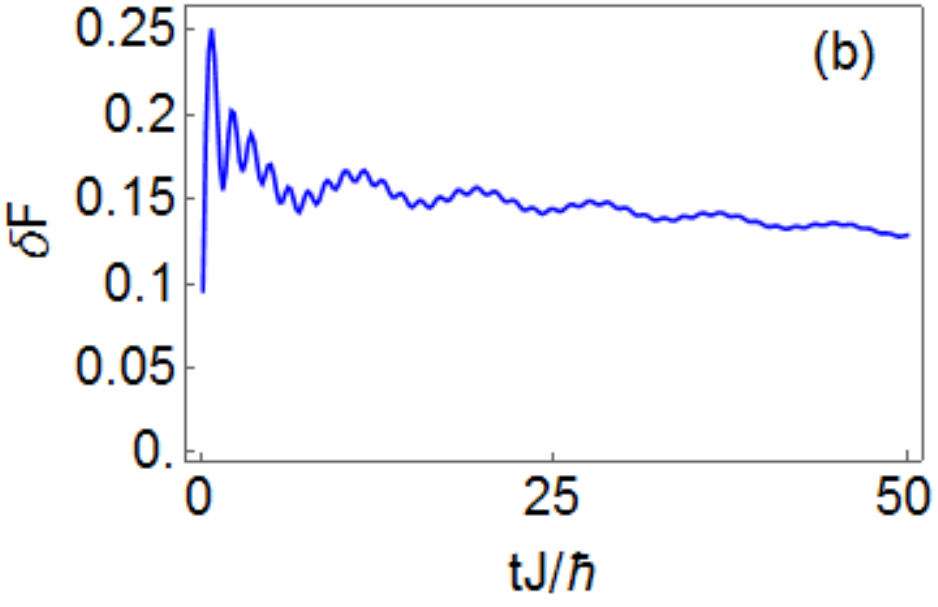}
\includegraphics[width=0.48\linewidth ,height=0.38\columnwidth]{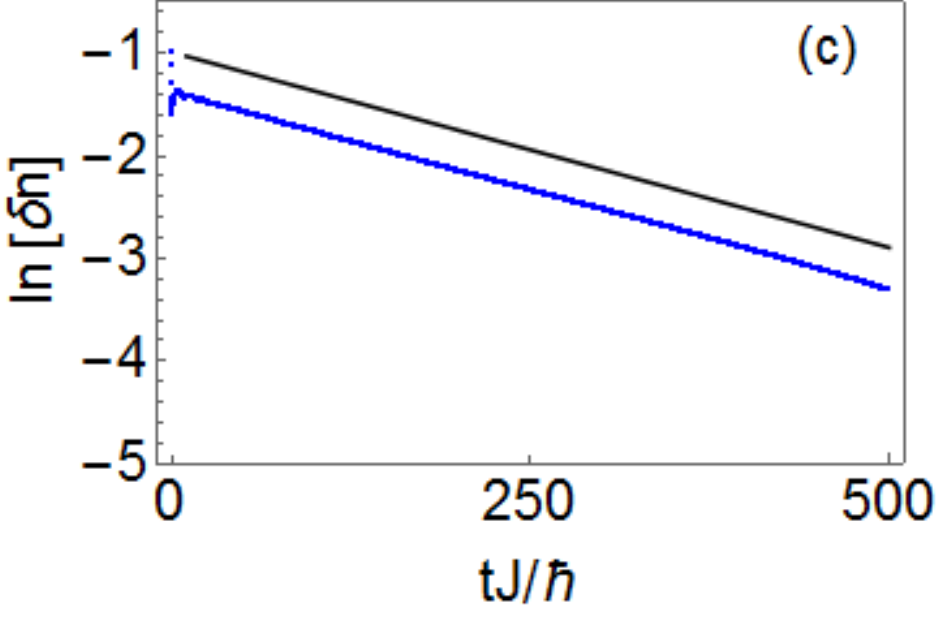}
\includegraphics[width=0.48\linewidth ,height=0.38\columnwidth]{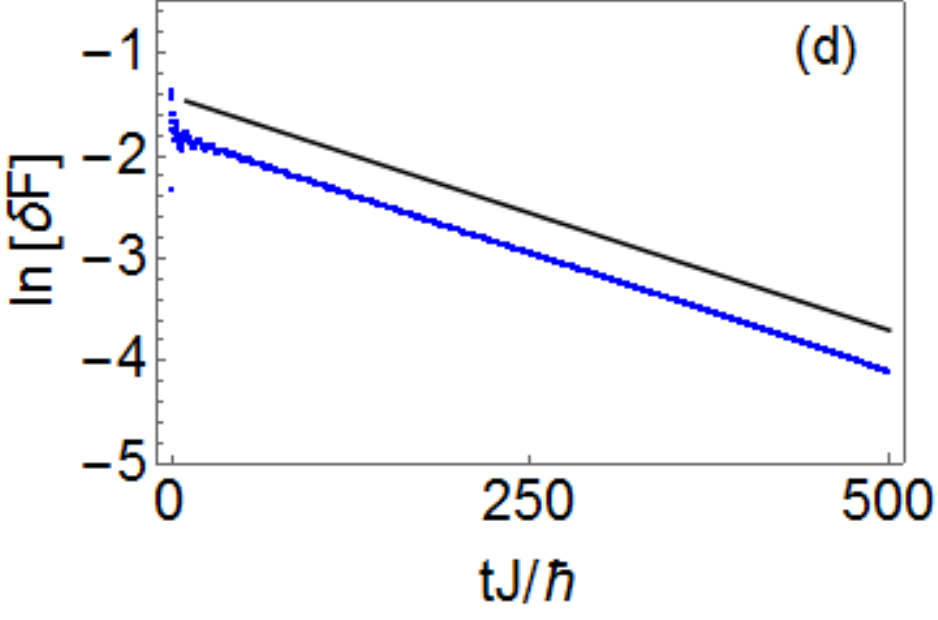}
\caption{(a) Evolution of $\delta n$ as a function of time $t$ (in
units of $\hbar/J$) for $g_i/J=2.5$, $g_f=0$, $\lambda/J=0.8$, and
$\hbar \omega_D= 10 \pi J$. (b) Similar plot for $\delta F$. (c)
Plot of $ \ln \delta n$ as a function of $t$ over longer time scale
showing the exponential decay of $\delta n$ with time. The black
line is the fit from which one obtains $\hbar \mu_n=0.0038 J$. (d)
Similar plot for $ \ln \delta F$ with $\hbar \mu_F=0.0044 J$. See
text for details.} \label{fig9}
\end{figure}

\begin{figure}[H]
\centering
\includegraphics[width=0.48\linewidth ,height=0.36\columnwidth]{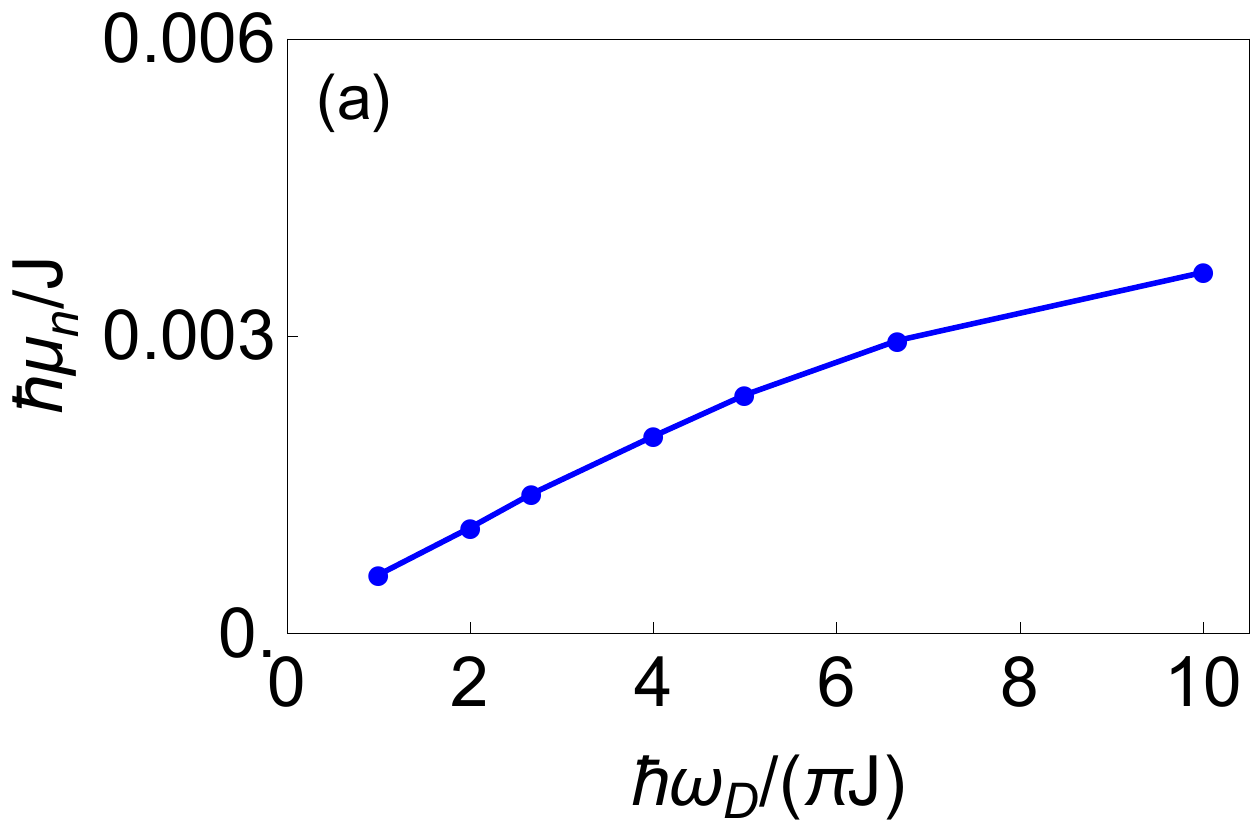}
\includegraphics[width=0.48\linewidth ,height=0.36\columnwidth]{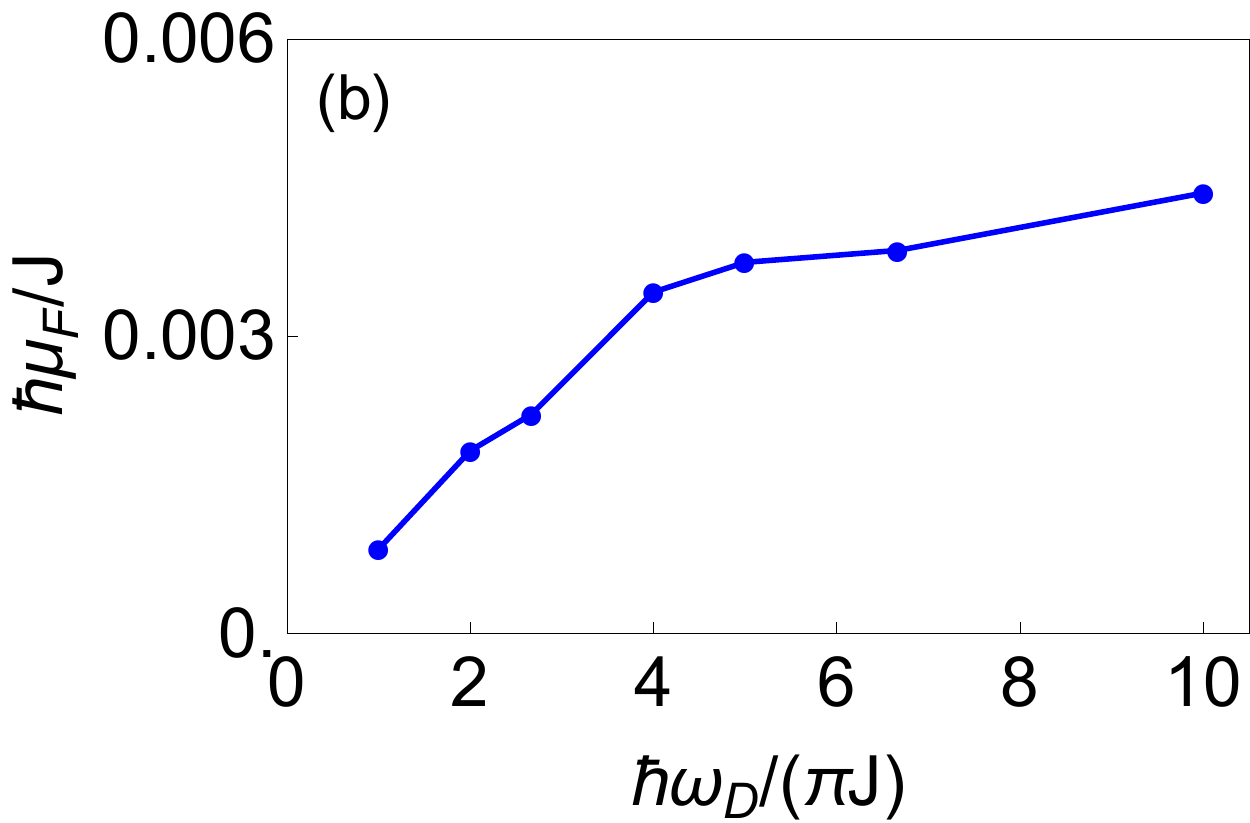}
\caption{(a) Plot of $\mu_n$ as a function of $\omega_D$ (in
units of $J/\hbar$ for $g_i/J=2.5$, $g_f=0$, $\lambda/J=0.8$.
(b) Similar plot for $\mu_F$. See text for details.}
\label{fig10}
\end{figure}

The result for this numerical study is shown in Fig.\ \ref{fig9} for
the Ising model in a transverse field. Fig.\ \ref{fig9}(a),(b) shows
the time variation the correlators $ \delta n= \sum_{k} (n_k -
n_k^{\rm steady\, state})$ and $ \delta F= \sum_k (F_k - F_k^{\rm
steady\, state})$ as a function of time $t$ (in units of $\hbar/J$).
We find that the correlators shows a decaying behavior which sets in
after brief oscillations for the first few cycles of the drive. The
nature of this decay is shown in Fig.\ \ref{fig9}(c),(d). We find
that, in contrast to fermionic bath, the presence of bosonic bath
leads to an exponential decay of the correlators to their steady
state value. The corresponding decay coefficients $\mu_n$ (of
$\delta n$) and $\mu_F$ (of $\delta F$) are plotted as a function of
the drive frequency $\omega_D$ in Fig.\ \ref{fig10}. This plot
indicates that $\mu_n$ and $\mu_F$ increases linearly with
$\omega_D$. This in turn implies that the decay of the correlators
as a function of number of drive cycles $n_0$ is independent of
$\omega_D$. Indeed, it is easy to see that if $\delta n (\delta F)
\sim \exp[-\mu'_{n(F)} n_0]$, then $\mu'_{n(F)} = \mu_{n(F)} T$ and
$\mu'_{n(F)}$ is thus independent of $\omega_D$ for $\mu_{n(F)} \sim
\omega_D$.

\begin{figure}[H]
\centering
\includegraphics[width=\linewidth]{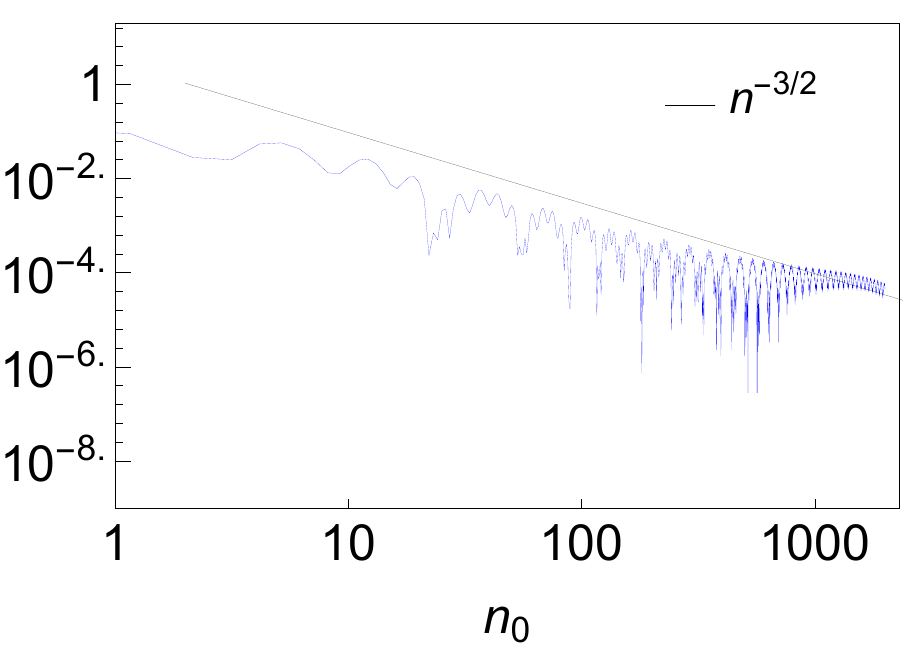}
\caption{Plot of $\delta n$ as a function of $n_0$ for $g_i/J=2.5$,
$g_f=0$, $\lambda/J=0.04$, and $\hbar \omega_D/J=10 \pi$. The plot
shows the change for power law to exponential decay around $n_c
\simeq 800$. See text for details.} \label{fig10o}
\end{figure}

Next, we investigate the role of the coupling parameter $\lambda$
behind such exponential decay of correlation functions. To this end,
we note that the exponential decay of the correlation functions sets
in at shorter time scales for larger $\lambda$; indeed it is
possible to define a critical number of drive cycles $n_c$ at any
drive frequency around which a crossover from power-law to
exponential decay takes place. This can be seen from the plot of $\delta
n$ as a function of $n_0$ in Fig.\ \ref{fig10o} where the crossover
from power-law to exponential behavior occurs around $n_0 \sim 800$.

\begin{figure}[H]
\centering
\includegraphics[width=0.49\linewidth ,height=0.34\columnwidth]{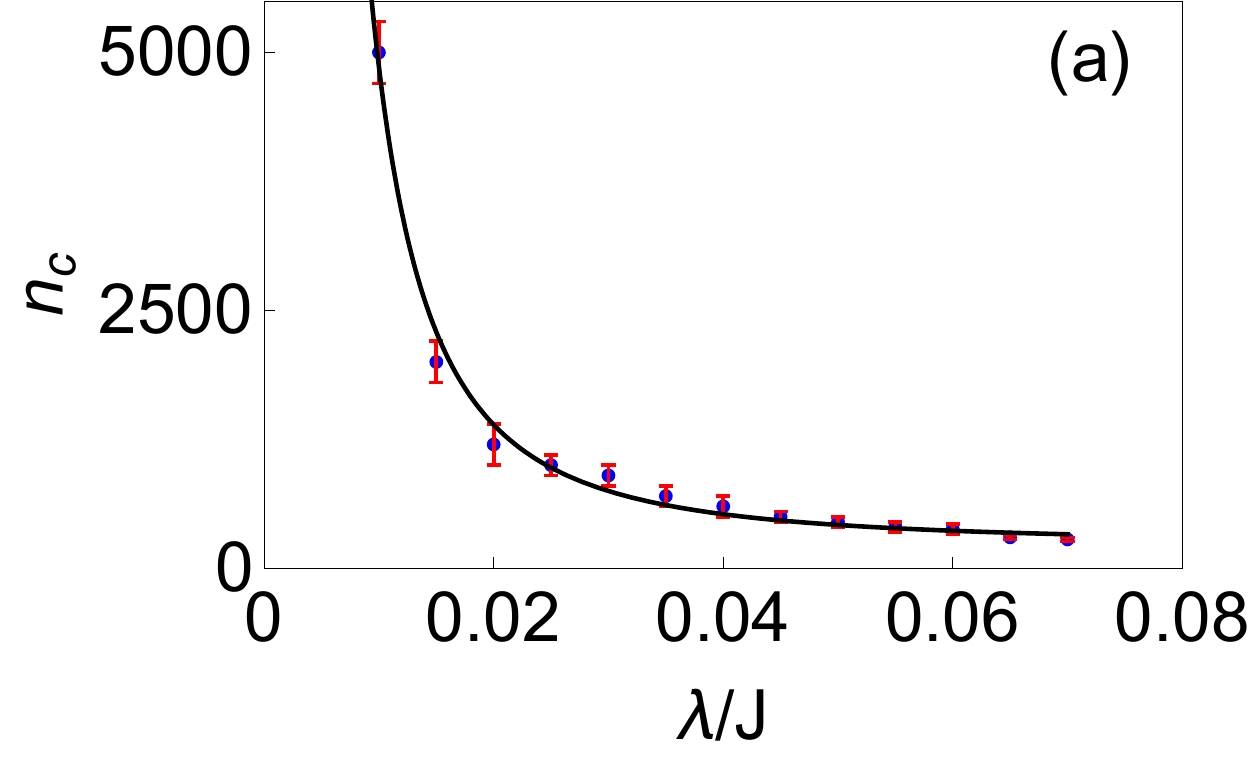}
\includegraphics[width=0.46\linewidth ,height=0.34\columnwidth]{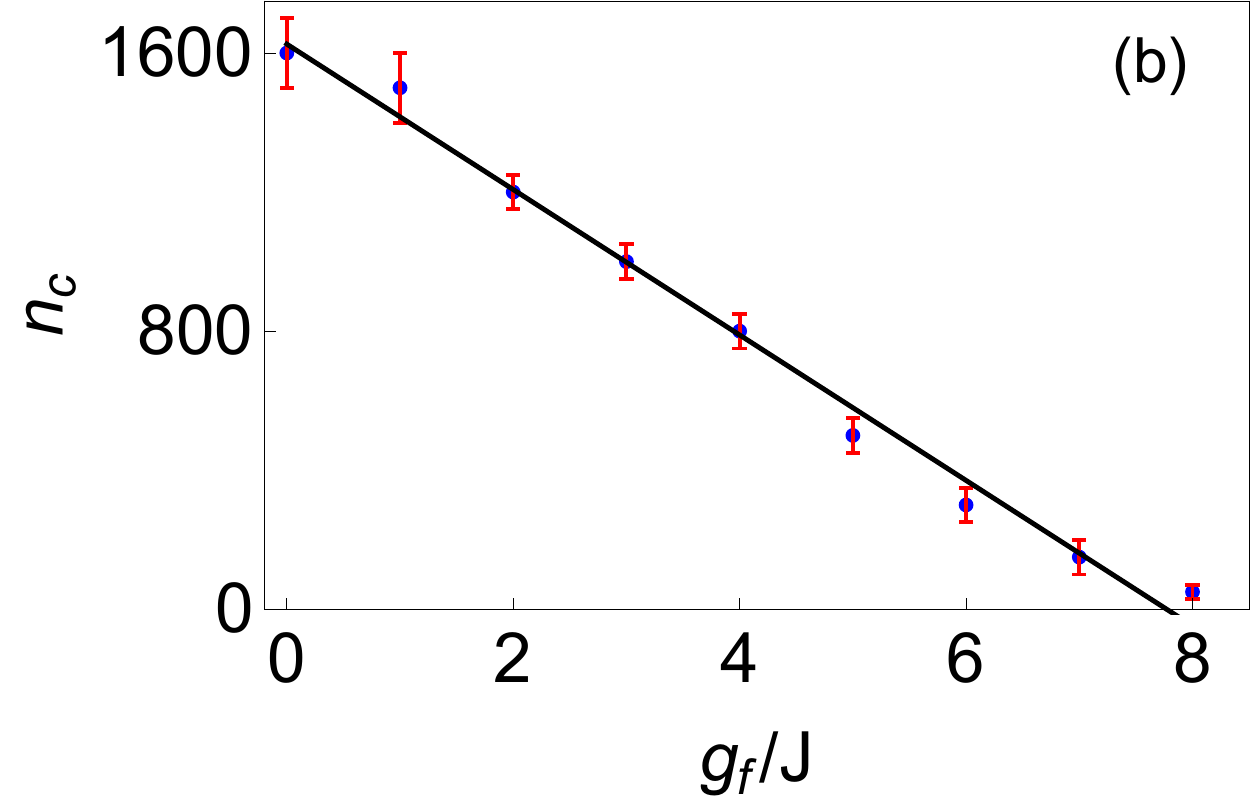}
\caption{(a) Plot of $n_c$ as a function of $\lambda$ (in
units of $J$) for $g_i/J=2.5$, $g_f=0$, $\hbar \omega_D/J=10 \pi$.
The lines shows a $1/\lambda^2$ fit to the data points indicated by
circles. (b) Plot of $n_c$ as a function of $g_f$ for
$\lambda=0.8 J$. All other parameters are same as in (a).
See text for details.} \label{fig11}
\end{figure}

For $n_0 \ll n_c$, the behavior of the system is analogous to a
closed Ising chain and the correlators display dynamical transition
as a function of frequency. For $n_0 \ge n_c$, the system shows the
exponential decay shown in Fig.\ \ref{fig10o}. A plot of $n_c$ as a
function of $\lambda/J$ is shown in Fig.\
\ref{fig11}(a); we find that $\delta n_c \sim 1/\lambda^2$. This
behavior can be understood as follows. We note that the
integrability of the Ising chain is destroyed by scattering between
different modes due to $H_1 \sim \lambda$; thus a simple Fermi
golden rule argument allows us to deduce that the time scale for
such scattering to become relevant would be $\sim 1/\lambda^2$. This
behavior is qualitatively similar to that of Fermi-Pasta-Ulam chain
\cite{fpuref} where it was shown that a finite strength of
integrability-breaking term is necessary to destroy the integrable
nature of the correlation functions. A plot of $n_c$ as a function
of $g_f/J$ for a fixed $\lambda/J$ is shown in
Fig.\ \ref{fig11}(b). The plot indicates that integrability breaking
behavior sets in more quickly for larger amplitude quenches. This
can be understood by considering the fact that larger amplitude
quenches amounts to larger energy transfer to the system which can
lead to quicker access to the bath degrees of freedom.

\begin{figure}[H]
\centering
\includegraphics[width=0.48\linewidth ,height=0.38\columnwidth]{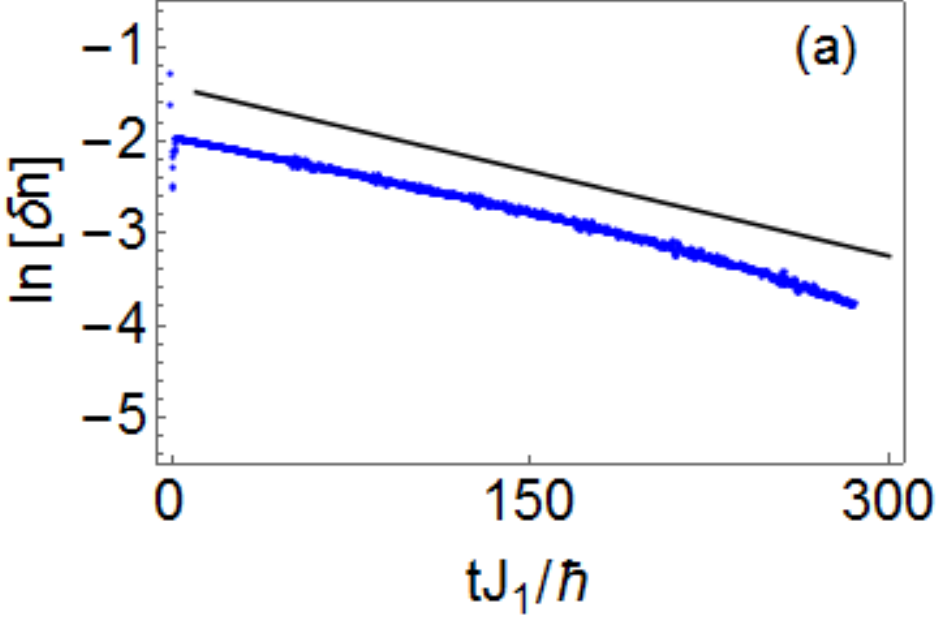}
\includegraphics[width=0.48\linewidth ,height=0.38\columnwidth]{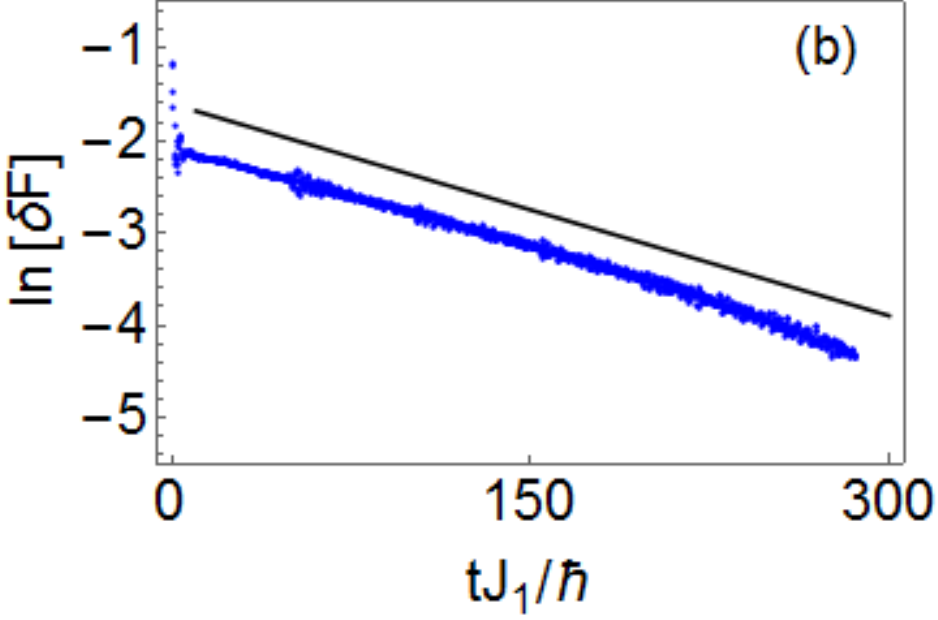}
\caption{(a) Plot of $ \ln \delta n$ as a function of $t$ (in units
of $\hbar/J_1$) for $J_{3f}/J_1=4$, $J_{3i}/J_1=5$,
$\lambda/J_1=0.8$, $\gamma_0/\lambda=0.2$, and $\hbar \omega_D= 10
\pi J_1$. The black line denotes the fit which yields $\hbar \mu_n^K
=0.0062 J_1$. (b) Similar plot for $ \ln \delta F$ with $\hbar
\mu_F^K =0.0072 J_1$. See text for details.} \label{fig12}
\end{figure}

Next we address the dynamics of the Kitaev model. For this, we scale
all quantities by $J_1$ and set $J_2=J_1$, $\hbar \omega_D=20 J_1$.
For all numerics, $J_3$ is varied using a square pulse protocol
between $J_{3i}=2.5J_1$ and $J_{3f}=0$ with a frequency $\omega_D$.
In Fig.\ \ref{fig12}, we show the dynamics of $\ln \delta n$ and
$\ln \delta F$ for the Kitaev model as a function of time $t$ (in
units of $\hbar/J_1$). The decay is again found to be exponential as
can be inferred from Fig.\ \ref{fig12}.

\begin{figure}[H]
\centering
\includegraphics[width=0.48\linewidth ,height=0.36\columnwidth]{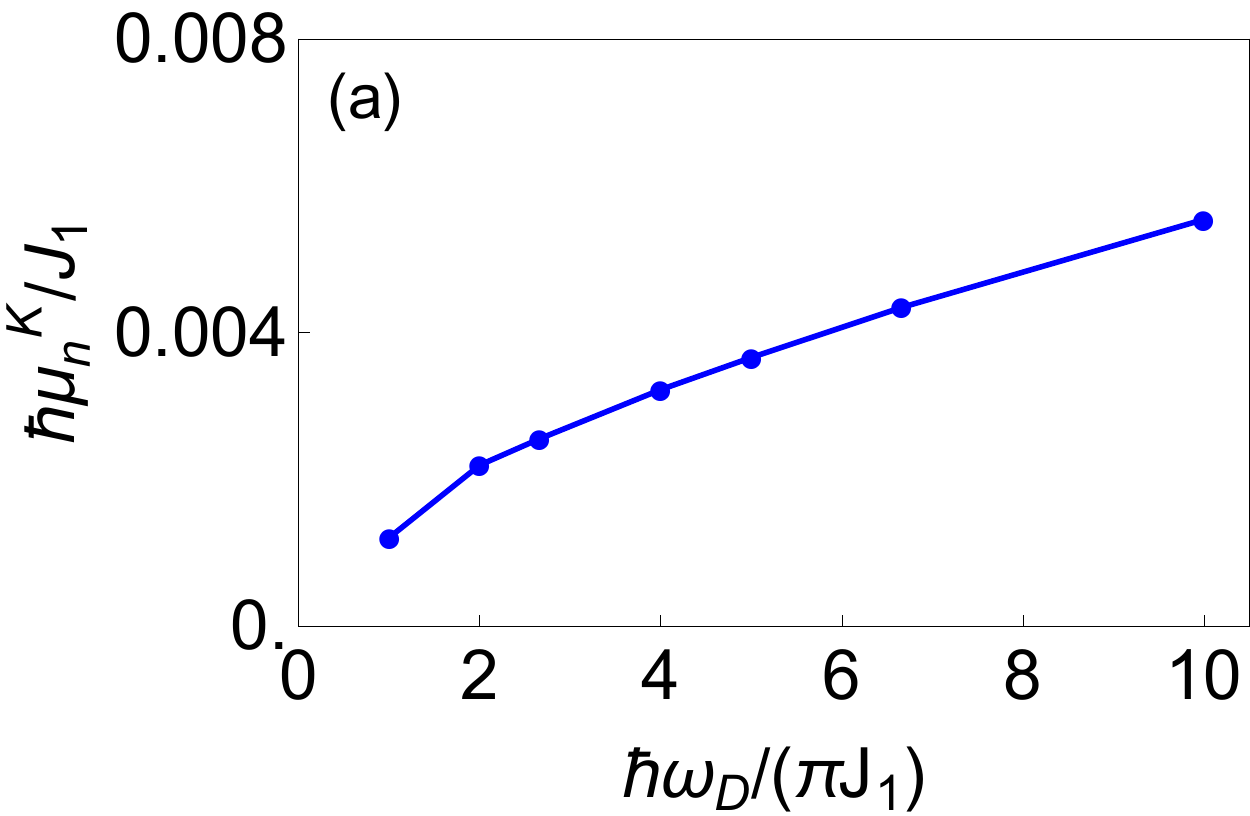}
\includegraphics[width=0.48\linewidth ,height=0.36\columnwidth]{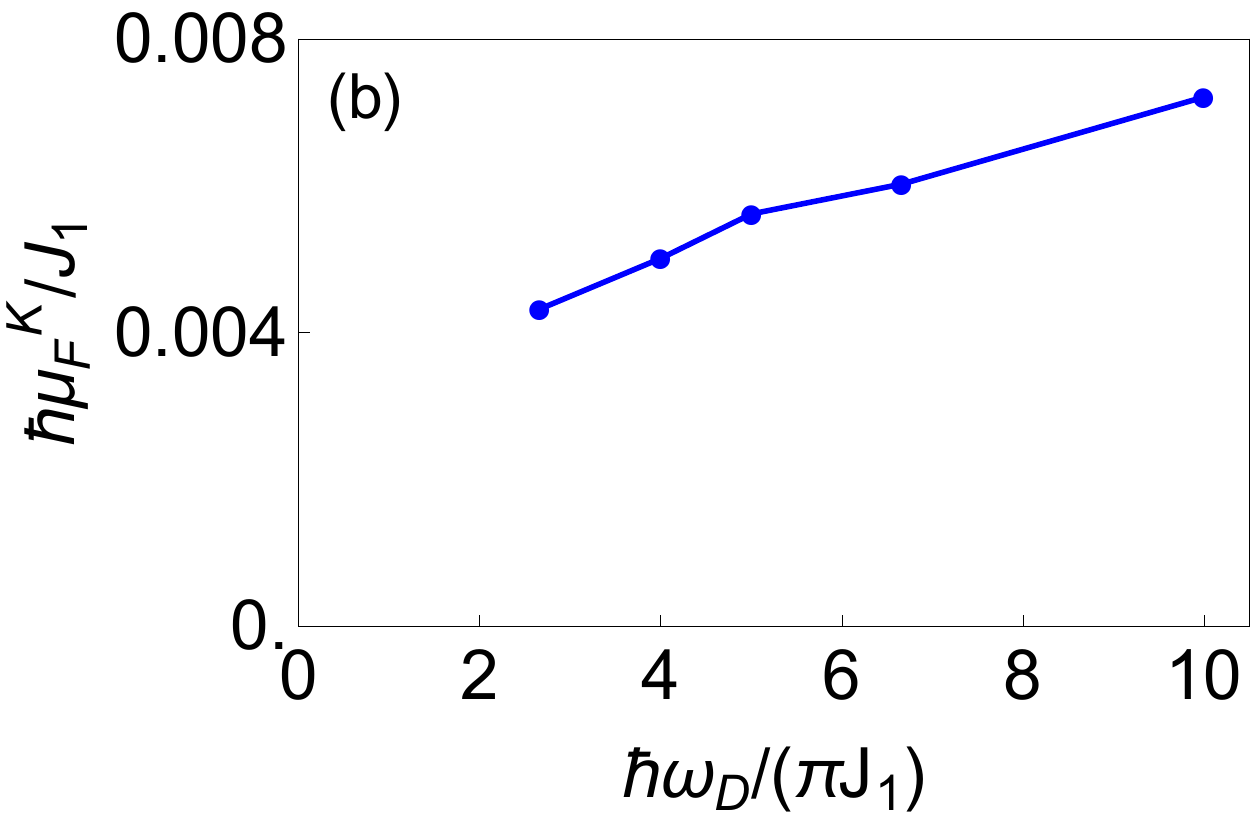}
\caption{(a) Plot of $\mu_n$ as a function of $\omega_D$ (in
units of $J_1/\hbar$) for $J_{3f}/J_1=4$, $J_{3i}/J_1=5$, and
$\lambda/J_1=0.8$. (b) Similar plot for $\mu_F$. All other
parameters are same as in Fig.\ \ref{fig12}. See text for details.}
\label{fig13}
\end{figure}

\begin{figure}[H]
\centering
\includegraphics[width=0.61\linewidth ,height=0.45\columnwidth]{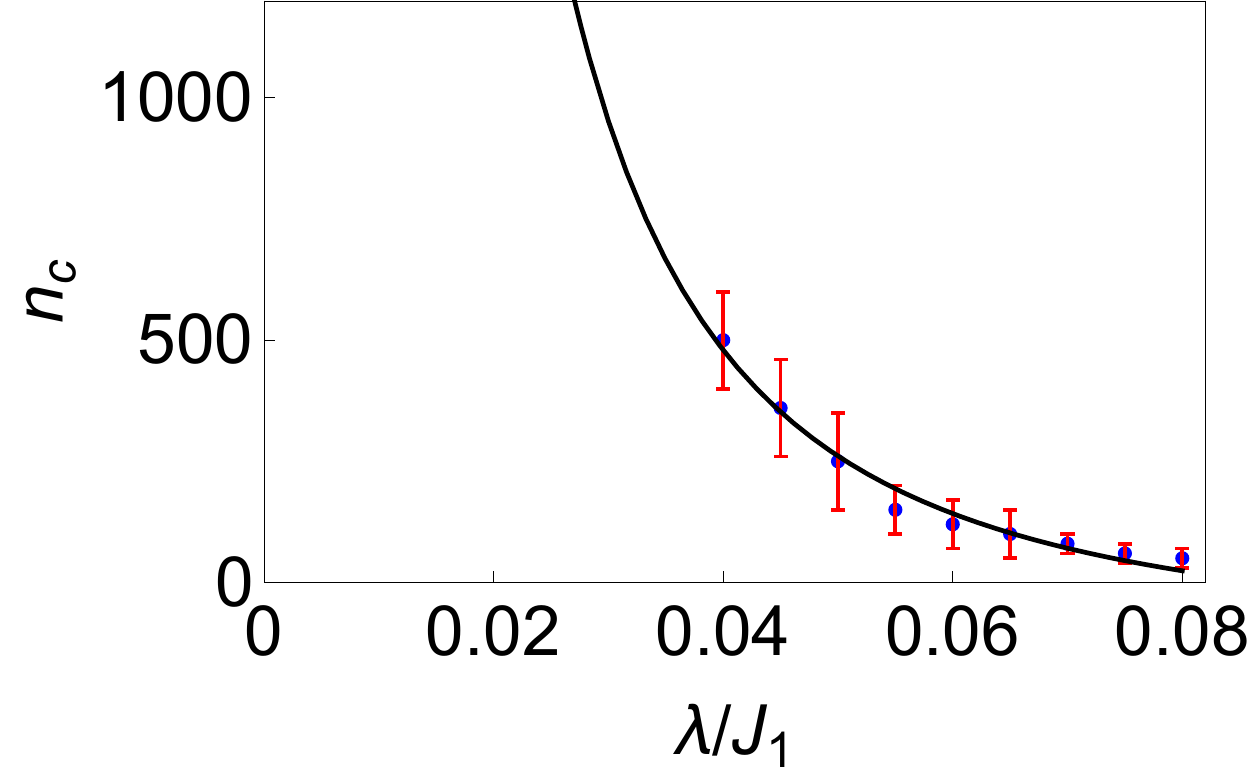}
\caption{ Plot of $n_c$ as a function of $\lambda$
(in units of $J_1$) for $J_{3f}/J_1=4$, $J_{3i}/J_1=5$, and $\hbar
\omega_D/J_1=10 \pi$. The dots represents data points where the
black line shows $1/\lambda^2$ fit to the data. All other parameters
are same as in Fig.\ \ref{fig12}. See text for details.}
\label{fig14}
\end{figure}

The decay coefficients of the correlation functions $\mu_n^K$ and
$\mu_F^K$ for the Kitaev model is shown in Fig.\ \ref{fig13}. These
plots indicates that both $\mu_n$ and $\mu_F$ for the Kitaev model
show an almost linear variation with drive frequency similar to
those for the Ising model. This in turn indicates that
$\mu^{'K}_{n,F}$ would be almost independent of $\omega_D$. The
variation of $n_c$ as a function of $\lambda$ shown in Fig.\
\ref{fig14} is also qualitatively similar to that for the 1D Ising
model. This seems to suggest that such behavior of $n_c$ is quite
general and one may expect to observe a dynamic transition for open
systems at sufficiently small $\lambda$; similar behavior is also
expected to be observed for CDW and superconducting systems.

\section{Discussion}
\label{diss}

In this work, we have studied the dynamics of a class of driven
integrable models coupled to an external bath. These models exhibit
drive frequency induced dynamical transitions in the absence of the
bath \cite{dt2,dt3}; our focus in this work has been to study the
fate of this dynamical transition in the presence of external baths.
Our study, which constitutes a generalization of such transition to
open quantum systems, reveals that the fate of such transitions
crucially depends on whether the bath breaks integrability of the
closed system.

For fermionic baths with linear coupling, where the integrability of
the closed system remains intact, we find that the transition
survives. For such baths, we provide a semi-analytic expression for
the Floquet eigenvalues corresponding to a square pulse drive
protocol. Using this, we chart out the different dynamical phases of
the system coupled to a fermionic bath. We demonstrate that the
coupling parameter between the system and the bath, $\lambda$, can
induce a new class of dynamical transitions which occur at large
$\lambda$. We note that such transitions occur at high drive
frequencies where the closed system exhibits $n_0^{-{d+1}/2}$ decay;
thus they do not have any analogue for closed integrable systems
studied earlier. In particular, we find transition lines in the
$\eta-\lambda$ plane for a fixed drive frequency; this demonstrates
the possibility of tuning these transition by varying fermionic bath
parameters.

In contrast, for bosonic baths which destroy the integrability of
the model, we use an equation of motion technique to the study the
dynamics. We restrict ourselves to the limit where the backreaction
of the system on the bath can be ignored. In this regime, we find
that all correlators decay to their steady state value
exponentially; these decays are characterized by decay coefficients
which varies linearly with the drive frequency for the 1D Ising model and 2D
Kitaev model. We note that such a decay sets in after a critical
number of drive cycles $n_c$; for $n_0 \ll n_c$, the power law decay
of the closed system survives. We chart out $n_c$ as a function of
the coupling strength $\lambda$ and show that $n_c \sim
1/\lambda^2$. This result indicates that for weak enough system-bath
coupling strength, one expects a large time window where the
dynamical transition would survive. We note that this result also
holds for weakly interacting closed fermion systems whose kinetic
term is given by $H_0$. This is seen by noting that our analysis for
the bosonic bath is carried out for $\hbar \omega_{\vec q}/J \gg1$;
in this regime integrating out the bath degrees of freedom leads to
a density-density interaction term for the fermions with strength $
\sim \lambda^2/(\hbar \omega_{\vec q_0})$.

Possible experimental platforms which can emulate such models
involves ultracold atom setups \cite{expt1,expt1b} or quantum dots
\cite{expt2,expt2b}. In particular, in Ref.\ \onlinecite{expt1b},
Dirac fermions described by $H_0$ was experimentally realized by
emulating fermions on a honeycomb lattice such as the one found in
graphene. The bosonic bath may be realized by coupling such a system
to  bath of oscillators; this was done for bosonic condensates
earlier \cite{expts3}. For fermionic bath, the setup in Ref.\
\onlinecite{expt1b} may be coupled to another 2D square lattice
which host fermions with tight binding dispersion. We propose
measurement of expectation of fermion density $n=\langle \sum_{\vec
k} \psi_{\vec k}^{\dagger} \psi_{\vec k}\rangle$ as a function of
time in such composite systems to verify the presence of two
different dynamical regimes.

In conclusion, we have studied driven dynamics of a class of
integrable fermionic models coupled to either fermionic or bosonic
baths. We have charted out the dynamical phases of these systems as
a function of drive frequency and system-bath parameters. Our
results show that the effect of these baths on the driven system
depends crucially on whether they preserve the integrability of
the system. We have discussed experiments which can test our
results.

\begin{acknowledgments}

The authors thank A. Sen and R. Ghosh for discussion.

\end{acknowledgments}

\end{document}